\documentclass[aps]{revtex4}
\usepackage{graphicx}
\DeclareGraphicsExtensions{.pdf,.eps}

\newcommand{\Tup}[0]{\mathbf \Gamma}
\newcommand{\Tep}[0]{\mathbf \Lambda}
\newcommand{\Tepmain}[0]{\Tep_{main}}
\newcommand{\Ttau}[0]{\mathbf{\Delta}}
\newcommand{\Tpi}[0]{\mathbf{\Pi}}
\newcommand{\TI}[0]{\mathbf I}

\newcommand{\mbf}[0]{\mathbf f}
\newcommand{\mbF}[0]{\mathbf F}
\newcommand{\bh}[0]{\mathbf h}
\newcommand{\br}[0]{\mathbf r}
\newcommand{\bT}[0]{\mathbf T}
\newcommand{\ba}{\mathbf a}
\newcommand{\bb}{\mathbf b}
\newcommand{\bc}{\mathbf c}
\newcommand{\bs}{\mathbf s}
\newcommand{\bz}{\mathbf z}

\newcommand{\lba}{a}
\newcommand{\lbb}{b}
\newcommand{\lbc}{c}
\newcommand{\lbh}{h}

\newcommand{\bsigma}{\mathbf \sigma}
\newcommand{\sh}[0]{\bsigma_{\lbh}}
\newcommand{\bhz}[0]{\mathbf h_{0}}
\newcommand{\bbS}{\mathbf S}

\newcommand{\etakh}{\eta_{i_k, \lbh}}
\newcommand{\etamh}{\eta_{i_m, \lbh}}
\newcommand{\etauh}{\eta_{i_\mu, \lbh}}
\newcommand{\etanh}{\eta_{i_1, \lbh}}

\begin{document}
\title{Extended Dynamical Equations of the Period Vectors
of Crystals under Constant External Stress to
Many-body Interactions}
\author{Gang Liu}
\email{gang.liu@queensu.ca,  gl.cell@outlook.com}
\affiliation{Centre for Advanced Computing,
Queen’s University,
99 University Avenue,
Kingston, Ontario, Canada K7L 3N6}

\date{August 08, 2018}



%
%
%
%
%


\begin{abstract}
Since crystals are made of periodic structures in space, predicting 
their three period
vectors starting from any values based on the inside 
interactions is a basic theoretical  physics problem.
For the general situation where crystals are under constant
external stress, we derived dynamical equations of the period vectors in
the framework of Newtonian dynamics, for pair potentials recently
(doi:/10.1139/cjp-2014-0518).
The derived dynamical equations show that the period
vectors are driven by the imbalance between the internal and external
stresses. 
This presents a physical process where when the external stress 
changes, the crystal structure changes accordingly, since the 
original internal stress can not balance the external stress. 
The internal stress has both
a full kinetic energy term and a full interaction term. It is 
extended to many-body interactions 
in this paper. As a result, all conclusions in
the pair-potential case also apply for many-body potentials.
\end{abstract}

\maketitle

\section{Introduction}

The spacial periodicity of the crystal structures is presented 
in almost all solid state physics 
books\cite{ktl,ibach,patterson}. Then a basic and 
general theory of predicting crystal 
structures under external pressure/stress is very desired. 
In 1980, by extending Andersen's idea\cite{a},
Parrinello and Rahman proposed their theory of such for the first
time in science history\cite{pr1,pr2}, when they met the same
problem in molecular dynamics (MD) simulations with the periodic
boundary condition being applied\cite{al,fre,haile}.  
Then many more efforts have been devoted to
this fundamental physics problem
\cite{ray1,nose,ryckaert,evans,ray2,ray3,hoover1985,
hoover1986,cll,wen,melchionna,jv,fo,martyna,ber,
szm,lww,lwssc,tuckerman,crystalpredictions,lgcjp1}. 
While all the rest of 
them were based on 
Lagrangian/Hamiltonian dynamics or minimizing (Gibbs) 
energy or enthalpy of the system, 
our recent effort\cite{lgcjp1} followed 
Newtonian dynamics.

According to the Born-Oppenheimer approximation, electrons
and ions of crystals are treated separately. Assuming the
motion of the electrons is always solved by applying quantum
mechanics with respect to any given configuration of the ions,
let us focus on the motion of the ions only, which is usually
described in the framework of classical physics. In other words,
electrons are regarded as a solvable media of interactions
among the ions, and in this paper all forces by
the electrons are assumed effectively 
included in the empirical many-body
interactions among ions. Then a crystal structure is reduced to
a periodic arrangement of exactly the same cells of ions in
three-dimensional space. 
As usually done in MD simulations, the spacial periodicity 
of the system structure is always 
assumed throughout this paper, 
however the ions move 
and the size and shape of the cells change. 
Using MD terms, the cell at the 
center of the crystal is called the MD cell and the ions in the 
MD cell are called the MD ions. 
Then the 
position vectors of the MD ions and the period vectors form the
complete degrees of freedom of the crystal. The period vectors
are also the edge vectors of a cell, which determine the size and
shape of cells. The period vectors may also be called basic vectors
or primitive translation vectors in solid state physics. Since crystals
are formed based on the interactions of the ions, their structures
should be predictable/determinable by dynamics. No doubt, the
dynamics of the MD ions is given by Newton's second law, then
the only task left is to derive the dynamics of the three independent
period vectors for crystals under external stress. 
All the dynamics should drive the system from a state 
of any positions 
of the MD ions and any size and shape of the MD cell 
towards  
an equilibrium 
state, where the structure is usually measured in  
experiments.


In our recent work\cite{lgcjp1}, while Newton's second law on
the MD ions was strictly preserved, the dynamical equations of
the period vectors were derived into the form where the period
vectors are driven by the imbalance between the internal and
external stresses, by repeatedly applying
Newton's laws. 
This means that when a crystal achieves an equilibrium state, 
the internal and external stresses must balance each other. 
It also presents a physical process where when the external stress 
changes, the crystal structure changes accordingly, because the 
original internal stress can not balance the external stress. 
Especially, the derived internal stress has both
the full kinetic energy term and the full interaction term. Since it
was done for pair-potential only and many-body interactions are
widely used\cite{aidan}, let us extend it to many-body
interactions here. As a result, all conclusions in the pair-potential case 
also apply for many-body potentials.

This paper is organized as follows, reflecting our three major steps.
After a description of our model in Sec.\ref{sec:Model},
Newton's second law is applied on half systems to get instantaneous
dynamical equations of the period vectors
in Sec.\ref{sec:Instancedynamics}.
Statistics of the above dynamical equations over indistinguishable
translated states is carried out to improve them
in Sec.\ref{sec:translatedstates}. Forces associated with momentum
transportation and statistics over ions' moving directions are further
implemented in the dynamical equations in
Sec.\ref{sec:transportmomentum}.
Sec.\ref{sec:Summary} is devoted to summary and discussion.

\section{Model}\label{sec:Model}

The limited macroscopic bulk of a crystal with an ``unlimited"
inside microscopic periodic structure is taken as the model.
We use $\ba$, $\bb$, and $\bc$ as the three independent period
vectors, forming a right-handed triad. Then each cell can be
denoted by the corresponding lattice translation vector
$\bT = T_\lba \ba + T_\lbb \bb + T_\lbc \bc$,
with integers $T_\lba$, $T_\lbb$, $T_\lbc$ ranging from negative
infinity to positive infinity.  As mentioned above,  the specific cell
of $\bT=0$ in the center is the MD cell, and the ions in it are the
MD ions. Since we study the properties of the inner part of the
bulk around the MD cell, far-away surface effects are neglected.

The external action on the surface is expressed by the constant
external stress tensor (or dyad) $\Tup$ .
The corresponding external forces are modeled as applied by
the surrounding external walls contacting the surface of the bulk.
For the case of constant external pressure $p$ ,
$\Tup=p\TI$,
where $\TI$ is an identity tensor or unit matrix, and the positive
direction is defined from inside to outside of the bulk.
By definition, the external force acting on an infinitesimal
surface area vector $d \bs$ of the bulk is $d\mbF = \Tup\cdot\, d\bs$.
The net external force on the bulk is
\begin{equation}
\mbF = \oint_{sf}\Tup\cdot\, d\bs =
\Tup\cdot\,\oint_{sf}d\bs=0,  \label{b02}
\end{equation}
where the integral is over all the surface of the bulk, and therefore
the bulk has no acceleration. The external stress $\Tup$ is assumed
to be symmetric, i.e., for all of its components $\Tup_{i,j}=\Tup_{j,i}$.
This  assumption ensures that the net external torque on the bulk is zero.

As said previously, the dynamics for the MD ions is always
Newton's second law
\begin{equation}
m_i \ddot \br_i = \mbF_i\ \ (i=1,2,\cdots ,n),\label{b03}
\end{equation}
where $\br_i$ is the position vector of the $i$th MD ion with
mass $m_i$, $\mbF_i$ is the net force acting on MD ion $i$
from all other ions of any cell (but no external force on MD ions
due to distance from the crystal surface), and $n$ is the total
number of MD ions. Then we will derive the dynamical equations
for the period vectors in the following.

For general purposes, consider $2$-body, $3$-body, $\cdots$,
up to $M$-body  interactions among any group of  ions in any
possible configurations. Since these many-body interactions are
independent on each other,
forces and potentials can be written as a summation of individual
$m$-body contributions. For example, the net force on MD ion $i$
can be expanded as
\begin{equation}
{\mbF}_{i}=\sum_{m=2}^M{\mbF}_{i}^{(m)}, \label{et01.000}
\end{equation}
where ${\mbF}_{{i}}^{(m)}$ is the contribution of $m$-body interactions.

For identifying an ion in the many-body interactions across the
whole crystal effectively, a simplified form of index $I_k$ was used
for it,  so that its position vector
can be expressed as
\begin{equation}
{\br}_{I_k}=I_{k,{\lba}}{\ba}+I_{k,{\lbb}}{\bb}+
I_{k,{\lbc}}{\bc}+{\br}_{i_k},
\label{et02}
\end{equation}
where $I_{k,{\lba}}$,  $I_{k,{\lbb}}$, and $I_{k,{\lbc}}$ are any
values of integers representing the cell in which it resides, and $i_k$,
ranging from 1 to $n$, refers to its corresponding image ion in the
MD cell.  This
means that $I_k$ represents the total four independent integer
variables of $(I_{k,{\lba}}$,  $I_{k,{\lbb}}$, $I_{k,{\lbc}}$, $i_k)$.
A summation over $I_k$ means the nested summations over the
four corresponding integers. As there are $m$ distinct ions
participating in any $m$-body interaction,
the subscript $k$ in $I_k$ is used to index the ions from 1
to $m$ in such an interaction.
Since no pair of ions can occupy the same physical location, for any
pair  of indexes $I_k$ and
 $I_{k^\prime}$, the expression
\begin{equation}
(I_{k,{\lba}}-I_{k^\prime, {\lba}})^2+
(I_{k,{\lbb}}-I_{k^\prime, {\lbb}})^2+
(I_{k,{\lbc}}-I_{k^\prime, {\lbc}})^2+
(i_k-i_{k^\prime})^2 \neq 0
\label{et02.2}
\end{equation}
is always assumed inside any $m$-body interaction throughout
this article.  This also means that for MD ion
 $i_{k^\prime}$
and any other ion $I_k$, the expression
$
(I_{k,{\lba}})^2+
(I_{k,{\lbb}})^2+
(I_{k,{\lbc}})^2+(i_{k}-i_{k^\prime})^2   \neq 0
$
is always true, and that for any two MD ions $i_{k}$ and
$i_{k^\prime}$, the mutual exclusive relationship  $i_{k} \neq i_
{k^\prime}$ is always true inside any $m$-body interaction.

Based on Newton's third law, the net force of the $m$-body
interaction in any given $m$-ion configuration should be zero
\begin{equation}
\sum_{k =1}^m{\mbf}_{I_k }^{(m)}({\br %
}_{I_1},{\br}_{I_2},{\br}_{I_3},\cdots ,{\br}_{I_m})=0,
\label{et02.3}
\end{equation}
where ${\mbf}_{I_k }^{(m)}({\br}_{I_1},{\br}_{I_2},{\br}_{I_3},
\cdots ,{\br}_{I_m})$ is the force acting on ion $I_k$ by all the
rest total $m-1$ ions.
Further considering the periodicity of the system, the net $m$-body
force acting on all MD ions should also be zero:
\begin{equation}
\sum_{i_1 =1}^n{\mbF}_{i_1}^{(m)}=0,  \label{et02.4}
\end{equation}
where
\begin{equation}
{\mbF}_{i_1}^{(m)}=
 {\frac {1} {(m-1)!}}
\sum_{\left \{ I_2,I_3,\cdots ,I_m  \right \} }
{\mbf}_{i_1}^{(m)}({\br %
}_{i_1},{\br}_{I_2},{\br}_{I_3},\cdots ,{\br}_{I_m}).
\label{et02.4.2}
\end{equation}
Equation (\ref{et02.4}) means no internal force can
push the system as a whole to accelerate.
With Eqs. (\ref{et01.000}) and (\ref{et02.4}) combined,
 it follows that the net of all forces acting on all MD ions
is zero, i.e.
\begin{equation}
\sum_{i=1}^nm_i\ddot {\br}_i=\sum_{i=1}^n{\mbF}_i=0,
\label{b12}
\end{equation}
where the summation indexes $i$ and $i_1$ are identical.
Employing the centre-of-mass coordinate system of the MD cell 
for all the work throughout this paper, the total momentum
of the MD cell is zero.

As the period vectors may change with time, the volume
$\Omega = \left(\ba\times\bb\right) \cdot \bc$ and shape
of the MD cell and those of the bulk should also change
accordingly.

\section{Instantaneous Dynamics}\label{sec:Instancedynamics}

\begin{figure}
  \begin{center}
    \hspace{-0.5cm}
    \includegraphics[width=0.7\textwidth]{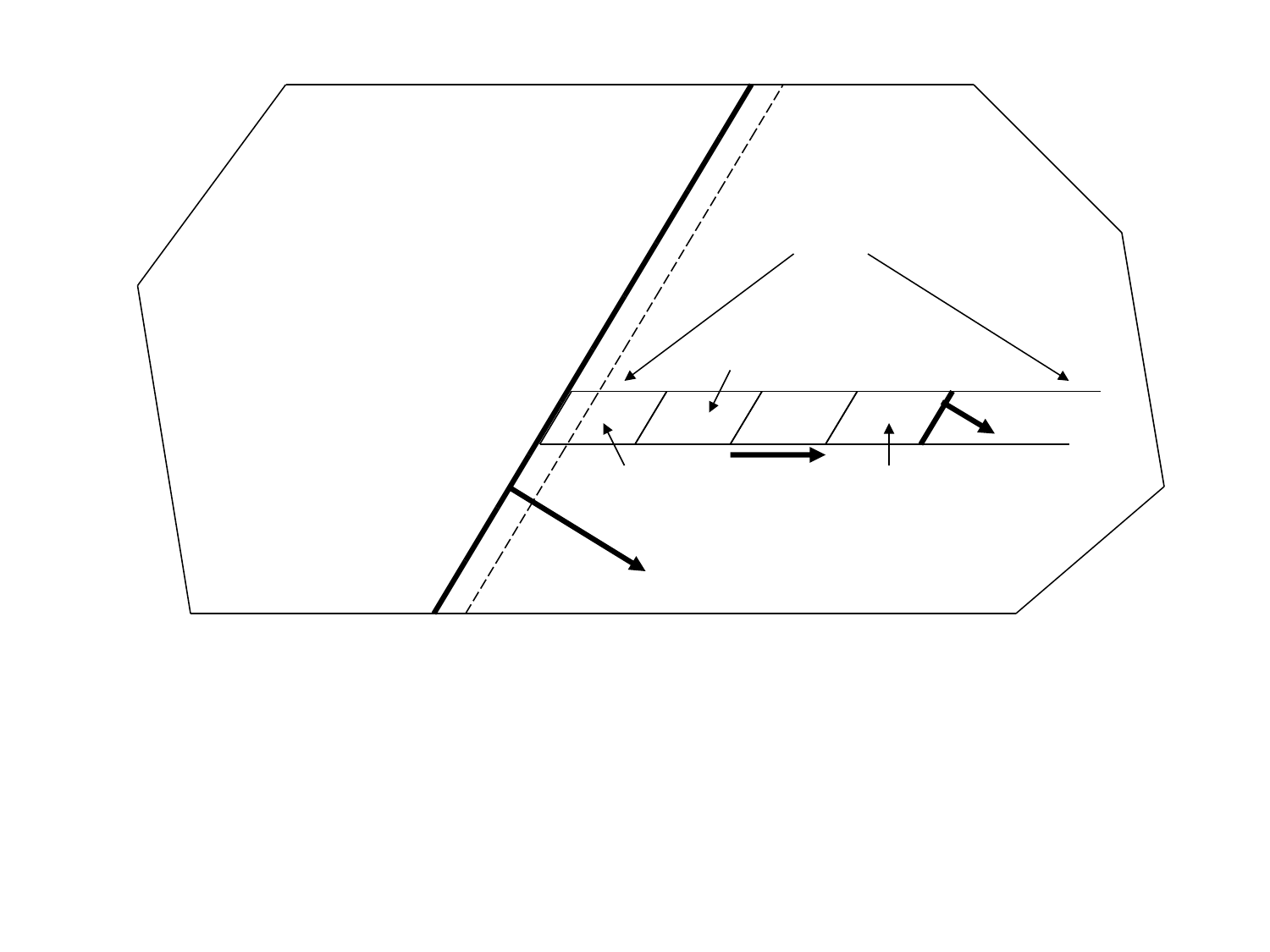}
  \end{center}
  \vspace{-9.5cm} \hspace{2.3cm} $P_{\lbh}$ \hspace{0.05cm} $Q_{\lbh}$ \\
  \vspace{1.2cm} \hspace{1.0cm} $L_{\lbh}$ \hspace{3.6cm}  $R_{\lbh}$ \\
  \vspace{0.0cm} \hspace{3.4cm} $B_{\lbh}$  \\
  \vspace{0.6cm} \hspace{1.3cm} cell ${\bh}$  \\
  \vspace{0.2cm} \hspace{6.8cm} ${\bsigma}_{\lbh}$  \\
  \vspace{0.4cm} \hspace{2.0cm} MD cell \hspace{0.4cm} ${\bh}$  \hspace{0.4cm} cell $3{\bh}$\\
  \vspace{0.4cm} \hspace{-1.8cm} ${\bbS}_{\lbh}$  \\
  \vspace{0.5cm} \hspace{-4.2cm} $P_{\lbh}^\prime$ \hspace{0.05cm} $Q_{\lbh}^\prime$ \\
  \vspace{7.3cm}
  \vspace{-0.6cm}
  \vspace{-0.1cm}
  \vspace{-0.3cm}
  \vspace{-0.2cm}
  \vspace{-0.2cm}
  \vspace{-0.1cm}
  \vspace{-0.3cm}

  \vspace{-5.0cm}

  \caption{
A sketch for the bulk of a crystal being cut by plane $P_{\lbh }P_{\lbh }^{\prime }$,
with a cross section area vector ${\bbS}_{\lbh}$. Plane $P_{\lbh}P_{\lbh}^{\prime }$ is
chosen such that for a given period vector
$\bh=\ba$, $\bb$, or $\bc$,
the right ($R_{\lbh}$) part contains all
$\bT = T_\lba \ba + T_\lbb \bb + T_\lbc \bc$
cells with $T_{\lbh}\geq 0$, and the left ($L_{\lbh}$) part contains all the
rest  ${\bT}$ cells with $T_{\lbh}< 0$.
The ``half-line-cell'' bar $B_{\lbh}$ is composed of the MD cell and cells $\bh$,
$2\bh$, $3\bh$, $4\bh$, etc., till the surface.
Newton's second law is applied to the $R_{\lbh}$ part for the dynamical equations
of the period vectors. (This figure was copied from \cite{lgcjp1}.)
}
  \label{fig1}
\end{figure}

In order to find the dynamical equations for the period vectors,
imagine a plane $P_{\lbh}P_{\lbh}^{\prime }$ that cuts the
model bulk into a right part and a left part,
with ${\bbS}_{\lbh}$ as the area vector of the cross section
between the two parts in the direction of pointing to the right
part, as shown in Fig.1. Plane $P_{\lbh}P_{\lbh}^{\prime }$
is chosen such that, for a given period vector
$\bh=\ba$, $\bb$, or $\bc$, the right ($R_\lbh$) part contains all
$\bT=T_{\lba}{\ba} +T_{\lbb}{\bb} +T_{\lbc}{\bc}$
cells with $T_{\lbh}\geq 0$, and the left ($L_{\lbh}$) part
contains all the
rest  ${\bT}$ cells with $T_{\lbh}< 0$.

Apply Newton's second law to a ``snapshot'' of the
right ($R_{\lbh}$) part.
Then, the net external force acting on the $R_{{\lbh}}$
part is
\begin{equation}
\mbF_{E,R} = \int_{R_{\lbh},sf}\Tup\cdot
d{\bs}=\Tup\cdot \int_{R_{\lbh},sf} d{\bs}=
\Tup\cdot {\bbS}_{\lbh},\label{b04}
\end{equation}
where the integral is over the surface of the bulk in the
$R_{\lbh}$ part. Let ${\mbF}_{L\rightarrow R}$ be the
net force acting on the $R_{\lbh}$ part by the $L_{\lbh}$
part. Then the dynamical equation of the $R_{\lbh}$ part is
\begin{equation}
M_R\ddot {\br}_{RC} = {\mbF}_{L\rightarrow R} +
\Tup\cdot {\bbS}_{\lbh},\label{b05}
\end{equation}
where $M_R$ is the total mass of the $R_{\lbh}$
part and $\ddot {\br}_{RC}$ is the acceleration of the centre of 
mass of the $R_{\lbh}$ part. 

Since surface effects are neglected,
${\mbF}_{L\rightarrow R}$ should be uniformly distributed
cell by cell across the section ${\bbS}_{\lbh}$ between the
two parts. Dividing Eq. (\ref{b05}) by
\begin{equation}
N_{\lbh}=\left| {\bbS}_{\lbh}\right| /\left|\sh\right|,\label{b07.0}
\end{equation}
where:
$\sh=\partial\Omega /\partial {\bh}$ is the (right) surface area
vector of a cell with respect to the period $\bh$, then
\begin{equation}
\frac 1{N_{\lbh}} M_R\ddot {\br}_{RC} = {\mbF}_{\lbh} +
\Tup\cdot \sh,\label{b07}
\end{equation}
where
\begin{equation}
{\mbF}_{\lbh} = \frac 1{N_{\lbh}}  {\mbF}_{L\rightarrow R},
\end{equation}
which is the net force, by the $L_{\lbh}$ part, acting on
the ``half-line-cell" bar $B_{\lbh}$ composed of the MD cell
and cells $\bh$,
$2\bh$, $3\bh$, $4\bh$, etc., till the surface, as shown in Fig.1.

Using Eq. (\ref{b12}), the left hand side of Eq. (\ref{b07}) becomes
\begin{equation}
\frac 1{N_{\lbh}} M_R\ddot {\br}_{RC} =
\frac 1{N_{\lbh}} \sum_{\bT\in R_{\lbh}}
\sum_{i=1}^nm_i\left(\ddot\br_i + \ddot\bT\right)  =
\frac{M_{cell}} {N_{\lbh}} \sum_{\bT\in R_{\lbh}} \ddot \bT,
\label{b13}
\end{equation}
where the total cell mass is $M_{cell}=\sum_{i=1}^nm_i$ and
the nested summations of $\sum_{\bT\in R_{\lbh}}\sum_{i=1}^n$
mean all ions in the $R_{\lbh}$ part are counted. Noticing that
$\ddot\bT=T_{\lba}\ddot\ba +T_{\lbb}\ddot\bb +T_{\lbc}\ddot\bc$,
Eq. (\ref{b13}) may be written as:
\begin{equation}
\frac 1{N_{\lbh}} M_R\ddot {\br}_{RC} =
\alpha _{\lbh,\lba}\ddot\ba +
\alpha _{\lbh,\lbb}\ddot\bb +
\alpha _{\lbh,\lbc}\ddot\bc,\label{b19}
\end{equation}
where
\begin{equation}
\alpha _{\lbh,\lbh^{\prime}}=
\frac{M_{cell}}{N_\lbh}
\sum_{\bT\in R_\lbh}
T_{{\lbh}^{\prime}}\ \ (\bh^{\prime} = \ba,\ \bb,\ \bc).\label{b20}
\end{equation}
In the $R_\lbh$ part,  $T_\lbh$ is always non-negative, but for any
 $T_{\lbh^{\prime}\neq\lbh}$, it is assumed there exists another
$-T_{\lbh^{\prime}}$ that cancels it in the above summation.
Therefore, all non-diagonal terms
$\alpha _{\lbh,\lbh^{\prime}\neq\lbh}$ are neglected.
Then Eq. (\ref{b07}) becomes
\begin{equation}
\alpha _{\lbh,\lbh} \ddot\bh =  {\mbF}_{\lbh} +
\Tup\cdot \sh.\label{b21}
\end{equation}

Considering all many-body interactions,
 the net force ${\mbF}_{{\lbh}}$ in  Eq. (\ref{b21})  can be written as:
\begin{equation}
{\mbF}_{{\lbh}}=\sum_{m=2}^M{\mbF}_{{\lbh}}^{(m)},  \label{et01}
\end{equation}
where ${\mbF}_{{\lbh}}^{(m)}$ is the contribution of $m$-body interactions.

The $m$-body interaction between the right and left part of the crystal
means that the participating ions must be distributed in both parts,
namely that not all
participating ions are in the same part.
Then ${\mbF}_{{\lbh}}^{(m)}$ is the net force on ions in the right
part of all such possible configurations divided by $N_{{\lbh}}$.
For total $t$
ions ($ m>t\geq1$)  in the right part (the rest of the ions are in
the left part at the same time), the corresponding net force for
all possibilities is
\begin{eqnarray}
{\mbF}_{t,{\lbh}}^{(m)}&=&\frac 1{N_{{\lbh}}}\frac 1{t!\left(
m-t\right) !}\sum_{\left\{ I_1,I_2,\cdots ,I_t\right\} }^{\left( I_{1,{\lbh}%
},I_{2,{\lbh}},\cdots ,I_{t,{\lbh}}\geq 0\right) }\sum_{\left\{
I_{t+1},I_{t+2},\cdots ,I_m\right\} }^{\left( I_{t+1,{\lbh}},I_{t+2,{\lbh}%
},\cdots ,I_{m,{\lbh}}<0\right) }\sum_{\mu =1}^t{\mbf}_{I_\mu }^{(m)}({\br %
}_{I_1},{\br}_{I_2},\cdots ,{\br}_{I_m}) \nonumber \\  \label{et03.1}
&=&
\frac 1{N_{{\lbh}}}\frac 1{t!\left(
m-t\right) !}\sum_{\left\{ I_1,I_2,\cdots ,I_t\right\} }^{\left( I_{1,{\lbh}%
},I_{2,{\lbh}},\cdots ,I_{t,{\lbh}}\geq 0\right) }\sum_{\left\{
I_{t+1},I_{t+2},\cdots ,I_m\right\} }^{\left( I_{t+1,{\lbh}},I_{t+2,{\lbh}%
},\cdots ,I_{m,{\lbh}}<0\right) }t{\mbf}_{I_1 }^{(m)}({\br %
}_{I_1},{\br}_{I_2},\cdots ,{\br}_{I_m}) \nonumber \\  \label{et03.2}
&=&
\frac 1{N_{{\lbh}}}\frac 1{(t-1)!\left(
m-t\right) !}\sum_{\left\{ I_1,I_2,\cdots ,I_t\right\} }^{\left( I_{1,{\lbh}%
},I_{2,{\lbh}},\cdots ,I_{t,{\lbh}}\geq 0\right) } 
\sum_{\left\{
I_{t+1},I_{t+2},\cdots ,I_m\right\} }^{\left( I_{t+1,{\lbh}},I_{t+2,{\lbh}%
},\cdots ,I_{m,{\lbh}}<0\right) } {\mbf}_{I_1 }^{(m)}({\br %
}_{I_1},{\br}_{I_2},\cdots ,{\br}_{I_m}),    \nonumber \\
&& \label{et03.3}
\end{eqnarray}
where $\sum_{\{\cdots \}}^{(\cdots )}$ denotes the nested summations over
indexes in $\{\cdots \}$ with conditions in $(\cdots )$, and commutability
among ions in each part is considered. If there are indexes listed in the
condition $(\cdots )$ expression, the condition applies to all of them,
otherwise applies to all the corresponding indexes listed
in $\{\cdots \}$. For example, $\sum_{\{i, j, k \}}^{(i, j<0)}$
restricts $i<0$ and $j<0$,
while $\sum_{\{i, j, k \}}^{\text{(positive)}}$
requires $i>0$, $j>0$, and $k>0$. 
If there is only one index in the condition expression, 
the brackets may be omitted. 
Throughout this article, all layers of nested summations should be realized
into reasonable forms even for special situations. For example, for $k=1$:
\begin{equation}
\sum_{\left\{ I_{2},I_{3},\cdots ,I_{k}   \right\} }^{\left( I_{2,{\lbh}%
},I_{3,{\lbh}},\cdots ,I_{k,{\lbh}}= 0\right) }
\sum_{\left\{
I_{k+1},I_{k+2},\cdots ,I_m\right\} }^{\left( I_{k+1,{\lbh}},I_{k+2,{\lbh}%
},\cdots ,I_{m,{\lbh}}>0\right) } \left( \cdots \right )
=
\sum_{\left\{
I_{k+1},I_{k+2},\cdots ,I_m\right\} }^{\left( I_{k+1,{\lbh}},I_{k+2,{\lbh}%
},\cdots ,I_{m,{\lbh}}>0\right) } \left( \cdots \right );
\end{equation}
while for $k=m$:
\begin{equation}
\sum_{\left\{ I_{2},I_{3},\cdots ,I_{k}   \right\} }^{\left( I_{2,{\lbh}%
},I_{3,{\lbh}},\cdots ,I_{k,{\lbh}}= 0\right) }
\sum_{\left\{
I_{k+1},I_{k+2},\cdots ,I_m\right\} }^{\left( I_{k+1,{\lbh}},I_{k+2,{\lbh}%
},\cdots ,I_{m,{\lbh}}>0\right) } \left( \cdots \right )
=
\sum_{\left\{ I_{2},I_{3},\cdots ,I_{k}   \right\} }^{\left( I_{2,{\lbh}%
},I_{3,{\lbh}},\cdots ,I_{k,{\lbh}}= 0\right) }
 \left( \cdots \right ).
\end{equation}

Remembering that
\begin{equation}
\sum_{I_1}^{I_{1,{\lbh}}\geq 0}\left ( \cdots \right ) =
\sum_{I_{1,{\lbh}}}^{I_{1,{\lbh}%
}\geq 0}\sum_{I_{1,{\lbh}^{\prime }}}
\sum_{I_{1,{\lbh}^{\prime \prime
}}}\sum_{i_1=1}^n \left ( \cdots \right ),   \label{et04}
\end{equation}
where ${\bh}^{\prime }$, ${\bh}^{\prime \prime }$
are also period vectors with possible values
$\left ( {\bh}, {\bh}^{\prime }, {\bh}^{\prime \prime } \right )
= \left ( {\ba}, {\bb}, {\bc} \right ) $,
or $ \left ( {\bb}, {\bc}, {\ba} \right ) $,
or $ \left ( {\bc}, {\ba}, {\bb} \right ) $ only,
considering
the crystal translatability,
employing $N_{{\lbh}%
}=\sum_{I_{1,{\lbh}^{\prime }}}\sum_{I_{1,{\lbh}^{\prime \prime }}}1$,
and setting $I_{1,{\lbh}^{\prime }}=I_{1,{\lbh}^{\prime \prime }}=0$,
Eq. (\ref{et03.3}) becomes 
\begin{eqnarray}
{\mbF}_{t,{\lbh}}^{(m)}&=&
\frac 1{(t-1)!\left( m-t\right) !}\sum_{I_{1,{\lbh}}=0}^{+\infty}
\sum_{i_1=1}^n
\sum_{\left\{ I_2,I_3,\cdots ,I_t  \right\} }^{\left( I_{2,{\lbh}},I_{3,%
{\lbh}},\cdots ,I_{t,{\lbh}}\geq 0\right) } 
\sum_{\left\{
I_{t+1},I_{t+2},\cdots ,I_m\right\} }^{\left( I_{t+1,{\lbh}},I_{t+2,{\lbh}%
},\cdots ,I_{m,{\lbh}}<0\right) } {\mbf}_{I_1}^{(m)}({\br}_{I_1},{\br}%
_{I_2},{\br}_{I_3},\cdots ,{\br}_{I_m}).  \label{et06}
\end{eqnarray}
Translating the system so that the cell
containing ion $I_1$, which is  
$I_{1,{\lbh}} {\bh} = 0 {\bh}$, $1 {\bh}$,
$2 {\bh}$, $3 {\bh}$, $\cdots$, 
becomes the MD cell, Eq. (\ref{et06})
can be further written as:
\begin{equation}
{\mbF}_{t,{\lbh}}^{(m)}=\frac 1{(t-1)!\left( m-t\right) !}
\sum_{l=0}^{-\infty}
\sum_{i_1=1}^n
\sum_{\left\{I_2,I_3,\cdots ,I_t \right\} }^{\left( I_{2,{\lbh}},I_{3,%
{\lbh}},\cdots ,I_{t,{\lbh}}\geq l\right) }\sum_{\left\{
I_{t+1},I_{t+2},\cdots ,I_m\right\} }^{\left( I_{t+1,{\lbh}},I_{t+2,{\lbh}%
},\cdots ,I_{m,{\lbh}}<l\right) }
{\mbf}_{i_1}^{(m)}({\br}_{i_1},{\br}_{I_2},%
{\br}_{I_3},\cdots ,{\br}_{I_m}),  \label{et08}
\end{equation}
and Eq. (\ref{et01}) becomes:
\begin{eqnarray}
{\mbF}_{{\lbh}} &=&\sum_{m=2}^M\sum_{t=1}^{m-1}{\mbF}_{t,{\lbh}%
}^{(m)}  \nonumber \\
&=&\sum_{m=2}^M\sum_{t=1}^{m-1}\frac 1{(t-1)!\left( m-t\right) !}
\sum_{l=0}^{-\infty}
\sum_{i_1=1}^n
\sum_{\left\{I_2,I_3,\cdots ,I_t  \right\} }^{\left( I_{2,{\lbh}},I_{3,%
{\lbh}},\cdots ,I_{t,{\lbh}}\geq l\right) } 
\sum_{\left\{
I_{t+1},I_{t+2},\cdots ,I_m\right\} }^{\left( I_{t+1,{\lbh}},I_{t+2,{\lbh}%
},\cdots ,I_{m,{\lbh}}<l\right) }
{\mbf}_{i_1}^{(m)}({\br}_{i_1},{\br}_{I_2},%
{\br}_{I_3},\cdots ,{\br}_{I_m}).  \label{et10}
\end{eqnarray}

Considering $m$-body potential
$\varphi ^{(m)}({\br}_{I_1},{\br}_{I_2},\cdots ,%
{\br}_{I_m})$, and supposing only $s$ $ (m \geq s \geq 1)$
of the $m$ ions are in the MD cell (all other ions are outside),
where only a fraction $s/m$ of the potential belongs to the cell,
the sum of all such potential belonging to the cell is:
\begin{eqnarray}
E_{p,cell,s}^{(m)}&=&\frac s{m}\frac 1{s!(m-s)!}
\sum_{\left\{i_1, i_2, \cdots, i_s \right\}}^{\text {(inside the cell)}}
\sum_{\left\{I_{s+1},I_{s+2},\cdots ,I_m  \right\}}^
{\text {(outside the cell)}}
\varphi ^{(m)}({\br}_{i_{1}},{\br}_{i_{2}},\cdots,  {\br}_{i_s},
{\br}_{I_{s+1}},{\br}_{I_{s+2}},\cdots ,{\br}_{I_m}).
\end{eqnarray}
Since the set of values $s^\prime = s-1 =0, 1, 2, \cdots, m-1$
means all possible situations where all ions, except ion $i_1$
(kept inside the cell), are placed inside or outside of the cell, one has:
\begin{eqnarray}
&&\frac 1{(m-1)!} \sum_{i_1}^{\text {(inside the cell)}}
\sum_{\left\{I_{2},I_{3},\cdots ,I_m  \right\}}
\varphi ^{(m)}({\br}_{i_{1}},
{\br}_{I_{2}},{\br}_{I_{3}},\cdots ,{\br}_{I_m})  \nonumber \\
&=&
\sum_{s^\prime=0}^{m-1}
\frac 1{s^\prime!(m-1-s^\prime)!}
\sum_{\left \{ i_1, i_2, \cdots, i_s \right \}}^
{\text {(inside the cell)}} 
\sum_{\left\{I_{s+1},I_{s+2},\cdots ,I_m  \right\}}^
{\text {(outside the cell)}}
\varphi ^{(m)}({\br}_{i_{1}},{\br}_{i_{2}},\cdots,  {\br}_{i_s},
{\br}_{I_{s+1}},{\br}_{I_{s+2}},\cdots ,{\br}_{I_m}).
\end{eqnarray}
Then the $m$-body cell potential energy becomes:
\begin{eqnarray}
E_{p,cell}^{(m)}
&=&
\sum_{s=1}^{m} E_{p,cell,s}^{(m)}
=\frac 1{m!} \sum_{i_1=1}^n \sum_{\left\{I_2,I_3,
\cdots ,I_m  \right\}
}\varphi ^{(m)}({\br}_{i_1},{\br}_{I_2},{\br}_{I_3},
\cdots ,{\br}_{I_m}).
\label{et22}
\end{eqnarray}
As a result, the total up to $M$-body cell potential energy is:
\begin{eqnarray}
E_{p,cell}&=&\sum_{m=2}^M
E_{p,cell}^{(m)}
=
\sum_{m=2}^M
\frac 1{m!} \sum_{i_1=1}^n \sum_{\left\{I_2,I_3,
\cdots ,I_m  \right\}
}\varphi ^{(m)}({\br}_{i_1},{\br}_{I_2},{\br}_{I_3},
\cdots ,{\br}_{I_m}).
\label{et21}
\end{eqnarray}

Making use of Eq. (\ref{et02}), take the derivative:
\begin{eqnarray}
-\frac \partial {\partial {\bh}}E_{p,cell}^{(m)}
&=&\frac 1{m !} \sum_{i_1=1}^n
\sum_{\left\{ I_2,I_3,\cdots ,I_m  \right\} } \sum_{k=2}^m
I_{k,{\lbh}}{\mbf%
}_{I_k}^{(m)}({\br}_{i_1},{\br}_{I_2},{\br}_{I_3},\cdots ,
{\br}_{I_m}) \nonumber \\
&=&\frac 1{m !} \sum_{i_1=1}^n  \sum_{\left\{ I_2,I_3,
\cdots ,I_m  \right\} } (m-1)
I_{m,{\lbh}}{\mbf%
}_{I_m}^{(m)}({\br}_{i_1},{\br}_{I_2},{\br}_{I_3},\cdots ,
{\br}_{I_m}) \nonumber \\
&=&\frac 1{m\left(
m-2\right) !} \sum_{i_1=1}^n  \sum_{\left\{ I_2,I_3,\cdots ,
I_m  \right\} }I_{m,{\lbh}}{\mbf%
}_{I_m}^{(m)}({\br}_{i_1},{\br}_{I_2},{\br}_{I_3},
\cdots ,{\br}_{I_m}),
\label{et25}
\end{eqnarray}
with the force:
\begin{equation}
{\mbf}_{I_k}^{(m)}({\br}_{I_1},{\br}_{I_2},\cdots ,{\br}%
_{I_m})=-\frac \partial {\partial {{\br}_{I_k}}} \varphi ^{(m)}({\br}_{I_1},{\br}%
_{I_2},\cdots ,{\br}_{I_m}).  \label{et26}
\end{equation}
The right side of Eq. (\ref{et25}) can be split into two
terms based on the
sign of $I_{m,{\lbh}}$ , so that:
\begin{equation}
-\frac \partial {\partial {\bh}}E_{p,cell}^{(m)}={\mbF}_{{\lbh,}%
+}^{(m)}+{\mbF}_{{\lbh,}-}^{(m)},  \label{et30}
\end{equation}
where:
\begin{eqnarray}
{\mbF}_{{\lbh,}+}^{(m)} &=&\frac 1{m\left( m-2\right) !}%
\sum_{i_1=1}^n \sum_{\left\{ I_2,I_3,\cdots ,I_m  \right\} }^
{I_{m,{\lbh}}>0}I_{m,{\lbh}}%
{\mbf}_{I_m}^{(m)}({\br}_{i_1},{\br}_{I_2},{\br}_{I_3},\cdots ,{\br}%
_{I_m})  \label{et41} \nonumber \\
 &=&\frac 1{m\left( m-2\right) !}
\sum_{l=0}^{+\infty }
\sum_{i_1=1}^n
\sum_{\left\{
I_2,I_3,\cdots ,I_m  \right\} }^{I_{m,{\lbh}}>l}
{\mbf}_{I_m}^{(m)}({\br}%
_{i_1},{\br}_{I_2},{\br}_{I_3},\cdots ,{\br}_{I_m}),  \label{et42}
\end{eqnarray}
and 
\begin{eqnarray}
{\mbF}_{{\lbh,}-}^{(m)} &=&\frac 1{m\left( m-2\right) !}%
\sum_{i_1=1}^n
\sum_{\left\{ I_2,I_3,\cdots ,I_m  \right\} }^{I_{m,{\lbh}}<0}
I_{m,{\lbh}}%
{\mbf}_{I_m}^{(m)}({\br}_{i_1},{\br}_{I_2},{\br}_{I_3},\cdots ,{\br}%
_{I_m}) \nonumber \\
&=&\frac{-1}{m\left( m-2\right) !}\sum_{l=0}^{-\infty}
\sum_{i_1=1}^n
\sum_{\left\{ I_2,I_3,\cdots ,I_m  \right\} }^{I_{m,{\lbh}}<l}
{\mbf}%
_{I_m}^{(m)}({\br}_{i_1},{\br}_{I_2},{\br}_{I_3},\cdots ,{\br}_{I_m}).
\label{et46}
\end{eqnarray}

By making use of the translatability to move the system so that the cell
$I_{m, \lbh} \bh + I_{m, \lbh^\prime} \bh^\prime +
I_{m, \lbh^{\prime\prime}} \bh^{\prime\prime} $, in which
ion $I_m$ resides, is translated to the MD cell, 
Eq. (\ref{et42}) becomes 
\begin{eqnarray}
{\mbF}_{{\lbh,}+}^{(m)}  
&=&\frac 1{m\left( m-2\right) !}
\sum_{l=0}^{+\infty }
\sum_{I_{1}}^{I_{1,{\lbh}} < -l}
\sum_{\left\{
I_2,I_3,\cdots ,I_{m-1}  \right\} }
\sum_{i_m=1}^n
{\mbf}_{i_m}^{(m)}({\br}%
_{I_1},{\br}_{I_2},{\br}_{I_3},\cdots ,{\br}_{i_m})
 \nonumber \\
&=&\frac 1{m\left( m-2\right) !}
\sum_{l=0}^{-\infty }
\sum_{I_{1}}^{I_{1,{\lbh}} < l}
\sum_{\left\{
I_2,I_3,\cdots ,I_{m-1}  \right\} }
\sum_{i_m=1}^n
{\mbf}_{i_m}^{(m)}({\br}%
_{I_1},{\br}_{I_2},{\br}_{I_3},\cdots ,{\br}_{i_m}).
\end{eqnarray}
Renaming ion $%
i_m$ as ion $I_1$ and ion $I_1$ as ion $i_m$, then
\begin{eqnarray}
{\mbF}_{{\lbh,}+}^{(m)}  
&=&\frac 1{m\left( m-2\right) !}
\sum_{l=0}^{-\infty }
\sum_{i_1=1}^n
\sum_{\left\{
I_2,I_3,\cdots ,I_{m-1}  \right\} }
\sum_{I_{m}}^{I_{m,{\lbh}} < l}
{\mbf}_{i_1}^{(m)}({\br}%
_{i_1},{\br}_{I_2},{\br}_{I_3},\cdots ,{\br}_{I_m}).
\end{eqnarray}
Expand it with respect to $t^{\prime }$ ,
the number of
ions distributed in the part of the crystal defined
by $I_{k,{\lbh}} \geq l$, of total $m-2$ ions indexed
from $I_2$ to $I_{m-1}$, then
\begin{eqnarray}
{\mbF}_{{\lbh,}+}^{(m)}
&=&
\frac 1m\sum_{t^{\prime
}=0}^{m-2}\sum_{l=0}^{-\infty }
\sum_{i_1=1}^n
\sum_{\left\{
I_2,I_3,\cdots ,I_{t^{\prime }+1}\right\} }^{\left( I_{2,{\lbh}%
},I_{3,{\lbh}},\cdots ,I_{t^{\prime }+1,{\lbh}}\geq l\right)
}\sum_{\left\{ I_{t^{\prime }+2},I_{t^{\prime }+3},\cdots ,I_{m-1}\right\}
}^{\left( I_{t^{\prime }+2,{\lbh}},I_{t^{\prime }+3,{\lbh}},\cdots ,I_{m-1,%
{\lbh}}<l\right) }
\sum_{I_{m}}^{I_{m,{\lbh}} < l}
\frac{{\mbf}_{i_1}^{(m)}({\br}_{i_1},{\br}_{I_2},{\br}_{I_3},%
\cdots ,{\br}_{I_m})}{t^{\prime }!\left( m-2-t^{\prime }\right) !}
\nonumber \\
&=&
\sum_{t=1}^{m-1}\frac{m-t}m{\mbF}_{t,{\lbh}}^{(m)},  \label{et56}
\end{eqnarray}
where $t$ is actually equal to $t^\prime +1$.

Now let us make use of the translatability to move the system so that cell
$I_{m, \lbh} \bh + I_{m, \lbh^\prime} \bh^\prime +
I_{m, \lbh^{\prime\prime}} \bh^{\prime\prime} $, in which
ion $I_m$ resides, is translated to the MD cell, 
then Eq. (\ref{et46}) becomes 
\begin{eqnarray}
{\mbF}_{{\lbh,}-}^{(m)} 
&=&\frac{-1}{m\left( m-2\right) !}\sum_{l=0}^{-\infty}
\sum_{I_{1}}^{I_{1,{\lbh}} > -l}
\sum_{\left\{
I_2,I_3,\cdots ,I_{m-1}  \right\} }
\sum_{i_m=1}^n
{\mbf}_{i_m}^{(m)}({\br}%
_{I_1},{\br}_{I_2},{\br}_{I_3},\cdots ,{\br}_{i_m}) 
 \nonumber \\
&=&\frac{-1}{m\left( m-2\right) !}\sum_{l=0}^{+\infty}
\sum_{I_{1}}^{I_{1,{\lbh}} > l}
\sum_{\left\{
I_2,I_3,\cdots ,I_{m-1}  \right\} }
\sum_{i_m=1}^n
{\mbf}_{i_m}^{(m)}({\br}%
_{I_1},{\br}_{I_2},{\br}_{I_3},\cdots ,{\br}_{i_m}).
\label{mr100}
\end{eqnarray}
Expand the above equation with 
respect to $t^{\prime }=t-1$ ,
the number of
ions distributed in the part of the crystal defined
by $I_{k,{\lbh}} > l$ , of total $m-2$ ions indexed
from $I_2$ to $I_{m-1}$, 
then it changes into
\begin{eqnarray}
{\mbF}_{{\lbh,}-}^{(m)} 
&=&\frac{-1}{m}
\sum_{t^{\prime}=0}^{m-2}
\sum_{l=0}^{+\infty}
\sum_{\left\{I_1,I_2,\cdots ,I_t \right\} }^{\left( I_{1,{\lbh}},I_{2,%
{\lbh}},\cdots ,I_{t,{\lbh}}> l\right) }\sum_{\left\{
I_{t+1},I_{t+2},\cdots ,I_{m-1}\right\} }
^{\left( I_{t+1,{\lbh}},I_{t+2,{\lbh}%
},\cdots ,I_{m-1,{\lbh}}\leq l\right) }
\sum_{i_m=1}^n
\frac {
{\mbf}_{i_m}^{(m)}({\br}%
_{I_1},{\br}_{I_2},{\br}_{I_3},\cdots ,{\br}_{i_m})
}{t^\prime!\left(m-2-t^\prime \right) !}.   \nonumber \\
&&
\label{mr105}
\end{eqnarray}

Meanwhile, employing Eq. (\ref{et02.3}), the first line of 
Eq. (\ref{et03.3}) can also be written as 
\begin{eqnarray}
{\mbF}_{t,{\lbh}}^{(m)}
&=&\frac {-1}{N_{{\lbh}}t!\left(
m-t\right) !}\sum_{\left\{ I_1,I_2,\cdots ,I_t\right\} }^{\left( I_{1,{\lbh}%
},I_{2,{\lbh}},\cdots ,I_{t,{\lbh}}\geq 0\right) }\sum_{\left\{
I_{t+1},I_{t+2},\cdots ,I_m\right\} }^{\left( I_{t+1,{\lbh}},I_{t+2,{\lbh}%
},\cdots ,I_{m,{\lbh}}<0\right) }\sum_{\mu =t+1}^m{\mbf}_{I_\mu }^{(m)}({\br %
}_{I_1},{\br}_{I_2},\cdots ,{\br}_{I_m}) \nonumber \\  
&=&\frac {-1}{N_{{\lbh}}t!\left(
m-t\right) !}\sum_{\left\{ I_1,I_2,\cdots ,I_t\right\} }^{\left( I_{1,{\lbh}%
},I_{2,{\lbh}},\cdots ,I_{t,{\lbh}}\geq 0\right) }\sum_{\left\{
I_{t+1},I_{t+2},\cdots ,I_m\right\} }^{\left( I_{t+1,{\lbh}},I_{t+2,{\lbh}%
},\cdots ,I_{m,{\lbh}}<0\right) }\left(
m-t\right){\mbf}_{I_m }^{(m)}({\br %
}_{I_1},{\br}_{I_2},\cdots ,{\br}_{I_m}) \nonumber \\  
&=&\frac {-1}{N_{{\lbh}}t!\left(
m-t-1\right) !}\sum_{\left\{ I_1,I_2,\cdots ,I_t\right\} }^{\left( I_{1,{\lbh}%
},I_{2,{\lbh}},\cdots ,I_{t,{\lbh}}\geq 0\right) }\sum_{\left\{
I_{t+1},I_{t+2},\cdots ,I_m\right\} }^{\left( I_{t+1,{\lbh}},I_{t+2,{\lbh}%
},\cdots ,I_{m,{\lbh}}<0\right) }
{\mbf}_{I_m }^{(m)}({\br %
}_{I_1},{\br}_{I_2},\cdots ,{\br}_{I_m}) \nonumber \\  
&=&\frac {-1}{t!\left(
m-t-1\right) !}\sum_{\left\{ I_1,I_2,\cdots ,I_t\right\} }^{\left( I_{1,{\lbh}%
},I_{2,{\lbh}},\cdots ,I_{t,{\lbh}}\geq 0\right) }\sum_{\left\{
I_{t+1},I_{t+2},\cdots ,I_{m-1}\right\} }
^{\left( I_{t+1,{\lbh}},I_{t+2,{\lbh}%
},\cdots ,I_{m-1,{\lbh}}<0\right) }
\sum_{I_{m,{\lbh}}}^{I_{m,{\lbh}}<0}
\sum_{i_m=1}^n
{\mbf}_{I_m }^{(m)}({\br %
}_{I_1},{\br}_{I_2},\cdots ,{\br}_{I_m}),      \label{mr001}
\end{eqnarray}
where in the last line, $I_{m,{\lbh}^{\prime }}
=I_{m,{\lbh}^{\prime \prime }}=0$, which means the cell
containing ion $I_m$ can be and only be 
$I_{m,{\lbh}} {\bh} = -1 {\bh}$, $-2 {\bh}$, 
$-3 {\bh}$, $\cdots$. Translating the system so that the cell
containing ion $I_m$ becomes the MD cell, Eq. (\ref{mr001})
becomes:
\begin{equation}
{\mbF}_{t,{\lbh}}^{(m)}
=\frac {-1}{t!\left(m-t-1\right) !}
\sum_{l=0}^{+\infty}
\sum_{\left\{I_1,I_2,\cdots ,I_t \right\} }^{\left( I_{1,{\lbh}},I_{2,%
{\lbh}},\cdots ,I_{t,{\lbh}}> l\right) }\sum_{\left\{
I_{t+1},I_{t+2},\cdots ,I_{m-1}\right\} }
^{\left( I_{t+1,{\lbh}},I_{t+2,{\lbh}%
},\cdots ,I_{m-1,{\lbh}}\leq l\right) }
\sum_{i_m=1}^n
{\mbf}_{i_m}^{(m)}({\br}_{I_1},{\br}_{I_2},%
{\br}_{I_3},\cdots ,{\br}_{i_m}).  \label{mr010}
\end{equation}

Combining Eqs. (\ref{mr105}) and Eq. (\ref{mr010}), 
then
\begin{equation}
{\mbF}_{{\lbh,}-}^{(m)}=\sum_{t=1}^{m-1}\frac tm{\mbF}_{t,{\lbh}%
}^{(m)}.  \label{et60}
\end{equation}
As a result
\begin{equation}
{\mbF}_{{\lbh}}=\sum_{m=2}^M\sum_{t=1}^{m-1}{\mbF}_{t,{\lbh}%
}^{(m)}=\sum_{m=2}^M\left( {\mbF}_{{\lbh,}+}^{(m)}+{\mbF}_{{\lbh,}%
-}^{(m)}\right)
=\sum_{m=2}^M-\frac \partial {\partial {\bh}}E_{p,cell}^{(m)}
=-\frac \partial {\partial {\bh}}E_{p,cell}.  \label{et61}
\end{equation}

Let us define the main interaction tensor for up to $M$-body
interactions as
\begin{equation}
\Tepmain=%
\frac{-1}\Omega \left[ \left( \frac{\partial E_{p,cell}}
{\partial {\ba}}%
\right) \otimes {\ba+}\left( \frac{\partial E_{p,cell}}
{\partial {\bb}}\right)
\otimes {\bb+}\left( \frac{\partial E_{p,cell}}{\partial {\bc}}\right)
\otimes {\bc}%
\right] =\sum_{m=2}^M
\Tepmain^{(m)}  \label{et62}
\end{equation}
with
\begin{equation}
\Tepmain^{(m)}=\frac{%
-1}\Omega \left[ \left( \frac{\partial E_{p,cell}^{(m)}}
{\partial {\ba}}%
\right) \otimes {\ba+}\left( \frac{\partial E_{p,cell}^{(m)}}
{\partial {\bb}}%
\right) \otimes {\bb+}\left( \frac{\partial E_{p,cell}^{(m)}}
{\partial {\bc}}%
\right) \otimes {\bc}\right] ,  \label{et62.0}
\end{equation}
then
\begin{equation}
{\mbF}_{{\lbh}}=\Tepmain\cdot \sh,  \label{et62.1}
\end{equation}
where  $\bh\cdot\sh=\Omega$ and
$\bh^{\prime}\cdot\sh=\bh^{\prime\prime}\cdot\sh=0$ are used.

Then Eq. (\ref{b21}) becomes
\begin{equation}
\alpha _{\lbh,\lbh} \ddot\bh =
\left( \Tepmain + \Tup\right ) \cdot \sh\ \
(\bh = \ba,\ \bb,\ \bc).\label{b21.9}
\end{equation}
The dynamical equation Eq. (\ref{b21.9}) is essentially the
same as in previous work\cite{lwssc}, for only constant external
pressure being considered.

\section{Microscopic Translated States} \label{sec:translatedstates}

\begin{figure}
  \begin{center}
    \hspace{-0.3cm}
    \includegraphics[width=0.7\textwidth]{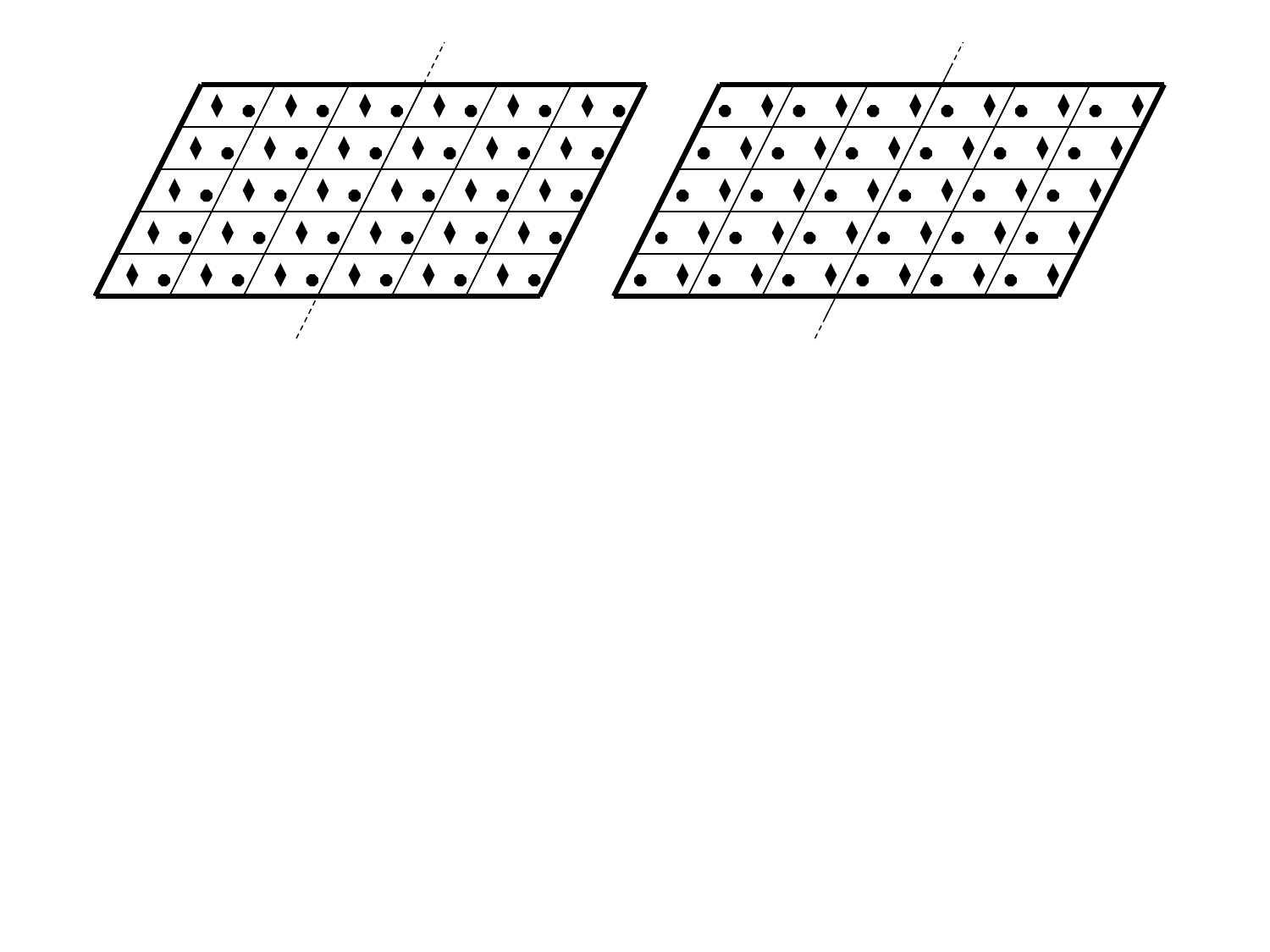}
  \end{center}
  \vspace{-9.4cm} \hspace{0.9cm} $P_{\lbh}$
  \vspace{-0.0cm} \hspace{5.2cm} $P_{\lbh}$ \\
  \vspace{2.2cm} \hspace{-1.5cm} $P_{\lbh}^{\prime}$
  \vspace{-0.0cm} \hspace{5.2cm} $P_{\lbh}^{\prime}$
  \vspace{7.3cm}
  \vspace{-1.7cm}
  \vspace{-5.5cm}
  \caption{
A sketch for two distinct states of the system
which are exactly the same in all microscopic details
except being translated slightly relative to each other.
As a set of image ions, the black diamonds are
on the right side in each cell of the right state, but
on the left side in each cell of the left state.
The black disks as another set of image ions are
on the other sides in the states.
In these states all right parts from plane $P_{\lbh}P_{\lbh}^{\prime }$
have the same number of cells.  (This figure was copied from \cite{lgcjp1}.)
}
  \label{fig2}
\end{figure}

As seen in Fig. 2, the two distinct states of the system are
exactly the same in all microscopic details except being translated
relative to each other. In these states all right parts from
plane $P_{\lbh}P_{\lbh}^{\prime }$ have the same number of cells
conceptually. Since they cannot be distinguished from a macroscopic
point of view, an unweighted average of  Eq. (\ref{b21.9}) or
Eq. (\ref{b21}) over all such configurations should be taken. Among
these, Eq. (\ref{b21}) is the same except the net force $\mbF_{\lbh}$,
acting on the half-line-cell bar $B_{\lbh}$ by the $L_{\lbh}$ part, as
in Fig. 1. Equivalently, this means that the
net force $\mbF_{\lbh}$ in Eq. (\ref{b21}) is replaced by the
unweighted average of it over all possible parallel locations of
cutting planes $P_{\lbh}P_{\lbh}^{\prime }$ that pass through
the MD cell in Fig. 1. For clarity, $P_{\lbh}P_{\lbh}^{\prime }$
was used in its original meaning (i.e., with fixed position),
but $Q_{\lbh}Q_{\lbh}^{\prime }$ was employed for such a
plane running from left to right.
For simplicity, $R_{\lbh}$ from Fig. 1 was separated into two
parts: $T_{\lbh}=0$ slab made of all cell of
$\bT=T_{\lba}\ba+T_{\lbb}\bb+T_{\lbc}\bc$ with $T_{\lbh}=0$,
and $R_{\lbh}^\prime$ part made of the rest of the cells. As an
example, for  $\bh=\ba$, $T_{\lbh}=0$ slab includes all cells of
$\bT=0\ba+T_{\lbb}\bb+T_{\lbc}\bc$ with $T_{\lbb}$
and $T_{\lbc}$ being any integers.
Since it makes a difference only if  $Q_{\lbh}Q_{\lbh}^{\prime }$
meets some ion(s) when it runs, we will only consider the situation
where there  are $s \geq 1$ ion(s) in the $T_{\lbh}=0$ slab
participating in the interactions. We will consider the following
three cases in sequence.

The first case is that there are other $t \geq 1$ ion(s) participating
in the $m$-body interaction appearing in the $L_{\lbh}$ part.
When the plane $Q_{\lbh}Q_{\lbh}^\prime$ runs from left to right
passing through the MD cell, the probability for MD ion  $i_k$
appearing on the left side of
$Q_{\lbh}Q_{\lbh}^\prime$ is
\begin{equation}
\etakh =
\frac {\left(\bh-\left(\br_{i_k}-\br_0\right)\right) \cdot\sh}{\Omega} =
\frac {\left(\bhz-\br_{i_k}\right) \cdot\sh}{\Omega},
\end{equation}
where
$\br_0$ is the position vector of the left-bottom and far-away
vertex of the MD cell and  $\bhz=\bh+\br_0$.
The following averaged net force acting on the $s$ ions by the rest ions
\begin{eqnarray}
{\mbF}_{c1,s,t,{\lbh}}^{(m)}&=&\frac 1{N_{{\lbh}}}
\frac 1{s!t!\left(m-t-s\right) !}
\sum_{\left\{ I_1,I_2,\cdots ,I_s\right\} }^{\left( I_{1,{\lbh}%
},I_{2,{\lbh}},\cdots ,I_{s,{\lbh}}= 0\right) }
\sum_{\left\{ I_{s+1},I_{s+2},\cdots ,I_{s+t}\right\} }^
{\left( I_{s+1,{\lbh}%
},I_{s+2,{\lbh}},\cdots ,I_{s+t,{\lbh}}< 0\right) }
\times \nonumber \\
&&
\sum_{\left\{
I_{s+t+1},I_{s+t+2},\cdots ,I_m\right\} }^
{\left( I_{s+t+1,{\lbh}},I_{s+t+2,{\lbh}%
},\cdots ,I_{m,{\lbh}}>0\right) }  \sum_{\mu =1}^s
\etauh
{\mbf}_{I_\mu }^{(m)}({\br %
}_{I_1},{\br}_{I_2},{\br}_{I_3},\cdots ,{\br}_{I_m}) \nonumber \\
&=&\frac 1{N_{{\lbh}}}
\frac 1{s!t!\left(m-t-s\right) !}
\sum_{\left\{ I_1,I_2,\cdots ,I_s\right\} }^{\left( I_{1,{\lbh}%
},I_{2,{\lbh}},\cdots ,I_{s,{\lbh}}= 0\right) }
\sum_{\left\{ I_{s+1},I_{s+2},\cdots ,I_{s+t}\right\} }^
{\left( I_{s+1,{\lbh}%
},I_{s+2,{\lbh}},\cdots ,I_{s+t,{\lbh}}< 0\right) }
\times   \nonumber \\
&&
\sum_{\left\{
I_{s+t+1},I_{s+t+2},\cdots ,I_m\right\} }^
{\left( I_{s+t+1,{\lbh}},I_{s+t+2,{\lbh}%
},\cdots ,I_{m,{\lbh}}>0\right) }  s
\etanh
{\mbf}_{I_1 }^{(m)}({\br %
}_{I_1},{\br}_{I_2},{\br}_{I_3},\cdots ,{\br}_{I_m}),
\label{et03.b.2}
\end{eqnarray}
should be excluded from $\mbF_{\lbh}$, as it was
unconditionally included in the $\mbF_{\lbh}$ previously.
The above equation can be simplified by translating the cell
where ion $I_1$ resides, in the $T_{\lbh}=0$ slab, to the MD cell
(then $I_1$ becomes $i_1$ and ${N_{{\lbh}}}$ can be reduced
to 1 by removing summations over $I_{1, \lbh{^\prime}}$
and $I_{1, \lbh{^{\prime \prime}}}$)
\begin{eqnarray}
{\mbF}_{c1,s,t,{\lbh}}^{(m)}&=&
\frac s{ s!t!\left(m-t-s\right) !}
\sum_{i_1=1}^n
\sum_{\left\{ I_2,I_3,\cdots ,I_s \right\} }^{\left( I_{2,{\lbh}%
},I_{3,{\lbh}},\cdots ,I_{s,{\lbh}}= 0\right) }
\sum_{\left\{ I_{s+1},I_{s+2},\cdots ,I_{s+t}\right\} }^
{\left( I_{s+1,{\lbh}%
},I_{s+2,{\lbh}},\cdots ,I_{s+t,{\lbh}}< 0\right) }
\times   \nonumber \\
&&
\sum_{\left\{
I_{s+t+1},I_{s+t+2},\cdots ,I_m\right\} }^
{\left( I_{s+t+1,{\lbh}},I_{s+t+2,{\lbh}%
},\cdots ,I_{m,{\lbh}}>0\right) }
\etanh
{\mbf}_{i_1 }^{(m)}({\br %
}_{i_1},{\br}_{I_2},{\br}_{I_3},\cdots ,{\br}_{I_m}) \nonumber \\
&=&
\frac s{ s!t!\left(m-t-s\right) !}
\sum_{i_1=1}^n
\sum_{\left\{ I_2,I_3,\cdots ,I_{t+1} \right\} }^
{\left( I_{2,{\lbh}%
},I_{3,{\lbh}},\cdots ,I_{t+1,{\lbh}}< 0\right) }
\sum_{\left\{ I_{t+2},I_{t+3},\cdots ,I_{t+s}\right\} }^
{\left( I_{t+2,{\lbh}%
},I_{t+3,{\lbh}},\cdots ,I_{t+s,{\lbh}}= 0\right) }
\times   \nonumber \\
&&
\sum_{\left\{
I_{s+t+1},I_{s+t+2},\cdots ,I_m\right\} }^
{\left( I_{s+t+1,{\lbh}},I_{s+t+2,{\lbh}%
},\cdots ,I_{m,{\lbh}}>0\right) }
\etanh
{\mbf}_{i_1 }^{(m)}({\br %
}_{i_1},{\br}_{I_2},{\br}_{I_3},\cdots ,{\br}_{I_m}),    \label{et03.b.3}
\end{eqnarray}
where ion index numbers were re-assigned. In the above expression,
there are $t \geq 1$ ion(s) in the $L_{\lbh}$ part. The situation
with $t =0$ ion(s) in the $L_{\lbh}$ part is considered in the other
two cases. There are $s \geq 1$ ion(s) in the $T_{\lbh}=0$ slab.
As ion $i_1$ is in the MD cell, there are additional $s-1$ ion(s) in
the $T_{\lbh}=0 $ slab. Actually the above expression is valid for
all situations with  $s^\prime = s-1 =0, 1, 2, \cdots,
m-t-1$ ions running anywhere in the $T_{\lbh}=0$
slab (except ${\br}_{i_1} $ position), and accordingly
$m-t-1-s^\prime $ ions running anywhere in
the $R_{\lbh}^\prime$ part. From total $m-t-1$ ions
numbered $t+2, t+3, \dots, m$ running anywhere in
the $R_{\lbh}$ part except ${\br}_{i_1} $ position,
considering all the possible situations placing $s^\prime$ ions
into the  $T_{\lbh}=0$ slab and putting all the rest ions in
the $R_{\lbh}^\prime$ part, the following equality emerges
\begin{eqnarray}
&&
\frac 1{ \left(m-t-1 \right) !}
\sum_{\left\{I_{t+2},I_{t+3},\cdots ,I_m \right\} }
^{\left(
 I_{t+2,{\lbh}},I_{t+3,{\lbh}},\cdots ,I_{m,{\lbh}}
\geq 0\right) }
\etanh
{\mbf}_{i_1 }^{(m)}({\br %
}_{i_1},{\br}_{I_2},{\br}_{I_3},\cdots ,{\br}_{I_m})   \nonumber \\
&&
=\sum_{s^\prime=0}^{m-t-1}
\frac{\etanh}
{ s^\prime !\left(m-t-1-s^\prime \right) !}
\sum_{\left\{ I_{t+2},I_{t+3},\cdots ,I_{t+s}
 \right\} }^{\left( I_{t+2,{\lbh}%
},I_{t+3,{\lbh}},\cdots ,I_{t+s,{\lbh}}= 0\right) } 
\sum_{\left\{
I_{t+s+1},I_{t+s+2},\cdots ,I_m\right\} }^
{\left( I_{t+s+1,{\lbh}},I_{t+s+2,{\lbh}%
},\cdots ,I_{m,{\lbh}}>0\right) }
{\mbf}_{i_1 }^{(m)}({\br %
}_{i_1},{\br}_{I_2},{\br}_{I_3},\cdots ,{\br}_{I_m})
.   \label{et03.b.4}
\end{eqnarray}

Combining Eq. (\ref{et03.b.3}) and Eq. (\ref{et03.b.4}),
\begin{eqnarray}
{\mbF}_{c1,t,{\lbh}}^{(m)}&=&
\sum_{s^\prime=0}^{m-t-1}
{\mbF}_{c1,s,t,{\lbh}}^{(m)}  \nonumber \\
&=&
\sum_{i_1=1}^n
\frac  {\etanh}
{ t!\left(m-t-1\right) !}
\sum_{\left\{ I_2,I_3,\cdots ,I_{t+1}\right\} }^
{\left( I_{2,{\lbh}%
},I_{3,{\lbh}},\cdots ,I_{t+1,{\lbh}}< 0\right) }
\sum_{\left\{
I_{t+2},I_{t+3},\cdots ,I_m \right\} }^
{\left( I_{t+2,{\lbh}},I_{t+3,{\lbh}%
},\cdots ,I_{m,{\lbh}} \geq 0\right) }
{\mbf}_{i_1 }^{(m)}({\br %
}_{i_1},{\br}_{I_2},{\br}_{I_3},\cdots ,{\br}_{I_m}) .      \nonumber \\
&&  \label{et03.b.5}
\end{eqnarray}

The second case is that all ions participating in the $m$-body
interaction reside in  $R_{\lbh}$ part, thus there is no ion in
the  $L_{\lbh}$ part, and there must be at least one ion in
the $T_{\lbh}=0$ slab and at least one ion in
the $R_{\lbh}^\prime$ part. The situation with no ion in
the $R_{\lbh}^\prime$ part will be considered in the last
case. Consider $s \geq 1$ ion(s) in the $T_{\lbh}=0$ slab
and the remaining total $m-s \geq 1$ ion(s) in
the $R_{\lbh}^\prime$ part.
For a given plane $Q_{\lbh}Q_{\lbh}^\prime$ cutting
the MD cell, the net force acting on the ions on the right
side of  $Q_{\lbh}Q_{\lbh}^\prime$ by those on the left
side should be added to $\mbF_{\lbh}$. Recalling
Eq. (\ref{et02.3}), equivalently the net force acting on the
ions on the left side of  $Q_{\lbh}Q_{\lbh}^\prime$ by
those on the right side should be subtracted
from $\mbF_{\lbh}$. With ion probabilities appearing on
the left side of  $Q_{\lbh}Q_{\lbh}^\prime$ considered,
the averaged net force on them can be written as
\begin{eqnarray}
{\mbF}_{c2,s,{\lbh}}^{(m)}&=&\frac 1{N_{{\lbh}}}
\frac 1{s!\left(m-s\right) !}
\sum_{\left\{ I_1,I_2,\cdots ,I_s\right\} }^{\left( I_{1,{\lbh}%
},I_{2,{\lbh}},\cdots ,I_{s,{\lbh}}= 0\right) }
\sum_{\left\{
I_{s+1},I_{s+2},\cdots ,I_m\right\} }^{\left( I_{s+1,{\lbh}},I_{s+2,{\lbh}%
},\cdots ,I_{m,{\lbh}}>0\right) } \sum_{\mu =1}^s
\etauh
{\mbf}_{I_\mu }^{(m)}({\br %
}_{I_1},{\br}_{I_2},{\br}_{I_3},\cdots ,{\br}_{I_m})
\nonumber \\
&=&\frac 1{N_{{\lbh}}}
\frac 1{s!\left(m-s\right) !}
\sum_{\left\{ I_1,I_2,\cdots ,I_s\right\} }^{\left( I_{1,{\lbh}%
},I_{2,{\lbh}},\cdots ,I_{s,{\lbh}}= 0\right) }
\sum_{\left\{
I_{s+1},I_{s+2},\cdots ,I_m\right\} }^{\left( I_{s+1,{\lbh}},
I_{s+2,{\lbh}%
},\cdots ,I_{m,{\lbh}}>0\right) } s
\etanh
{\mbf}_{I_1 }^{(m)}({\br %
}_{I_1},{\br}_{I_2},{\br}_{I_3},\cdots ,{\br}_{I_m})
\nonumber \\  \label{et03.c.2}
&=&
\frac 1{(s-1)!\left(m-s\right) !}
\sum_{i_1=1}^{n}
\sum_{\left\{ I_2,I_3,\cdots ,I_s \right\} }^{\left( I_{2,{\lbh}%
},I_{3,{\lbh}},\cdots ,I_{s,{\lbh}}= 0\right) }
\sum_{\left\{
I_{s+1},I_{s+2},\cdots ,I_m\right\} }^{\left( I_{s+1,{\lbh}},
I_{s+2,{\lbh}%
},\cdots ,I_{m,{\lbh}}>0\right) }
\etanh
{\mbf}_{i_1 }^{(m)}({\br %
}_{i_1},{\br}_{I_2},{\br}_{I_3},\cdots ,{\br}_{I_m}) .  \label{et03.c.3}
\end{eqnarray}

The last case is that all ions participating in
the $m$-body interaction are in the $T_{\lbh}=0$ slab.
For a given plane $Q_{\lbh}Q_{\lbh}^\prime$
cutting the MD cell, the net force acting on the ions
on the right side of  $Q_{\lbh}Q_{\lbh}^\prime$ by
those on the left side should be added to $\mbF_{\lbh}$.
Based on Eq. (\ref{et02.3}), equivalently the net force
acting on the ions on the left side
of  $Q_{\lbh}Q_{\lbh}^\prime$ by those on the right
side should be subtracted from $\mbF_{\lbh}$. As a
matter of fact, only when ions are distributed on both
sides of plane $Q_{\lbh}Q_{\lbh}^\prime$, such forces
should be considered. For a given configuration of
the $m$ ions, assuming ion $I_m$ is the last one to
be crossed by plane $Q_{\lbh}Q_{\lbh}^\prime$ when
it runs from left to right,  the probability for ion $I_k$
appearing on the left side of plane
$Q_{\lbh}Q_{\lbh}^\prime$ and ion $I_m$ on the right
side  is   $\left(\br_{i_m}-\br_{i_k}\right) \cdot\sh/\Omega$,
then the averaged net force of the given configuration to
be subtracted from $\mbF_{\lbh}$ is
\begin{eqnarray}
{\mbf}_{c3,{\lbh}}^{(m)}&=&
\sum_{\mu =1}^m
\frac {\left(\br_{i_m}-\br_{i_\mu} \right) \cdot \sh} {\Omega}
{\mbf}_{I_\mu }^{(m)}({\br %
}_{I_1},{\br}_{I_2},{\br}_{I_3},\cdots ,{\br}_{I_m})
 \nonumber \\
&=&
\sum_{\mu =1}^m
\frac {\left(\left(\bhz-\br_{i_\mu}\right) -
\left(\bhz-\br_{i_m} \right)  \right) \cdot \sh} {\Omega}
{\mbf}_{I_\mu }^{(m)}({\br %
}_{I_1},{\br}_{I_2},{\br}_{I_3},\cdots ,{\br}_{I_m})
\nonumber \\
&=&
\sum_{\mu =1}^m
\etauh
{\mbf}_{I_\mu }^{(m)}({\br %
}_{I_1},{\br}_{I_2},{\br}_{I_3},\cdots ,{\br}_{I_m}) -
\etamh
\sum_{\mu =1}^m
{\mbf}_{I_\mu }^{(m)}({\br %
}_{I_1},{\br}_{I_2},{\br}_{I_3},\cdots ,{\br}_{I_m}).\label{et03.c.4}
\end{eqnarray}
Applying Eq. (\ref{et02.3}), the last term in the last
equation becomes zero, then
the averaged net force of all such configurations is
\begin{eqnarray}
{\mbF}_{c3,{\lbh}}^{(m)}&=&\frac 1{N_{{\lbh}}}
\frac 1{m!}
\sum_{\left\{ I_1,I_2,\cdots ,I_m\right\} }^{\left( I_{1,{\lbh}%
},I_{2,{\lbh}},\cdots ,I_{m,{\lbh}}= 0\right) }
{\mbf}_{c3,{\lbh}}^{(m)} \nonumber \\
&=&\frac 1{N_{{\lbh}}}
\frac 1{m!}
\sum_{\left\{ I_1,I_2,\cdots ,I_m\right\} }^{\left( I_{1,{\lbh}%
},I_{2,{\lbh}},\cdots ,I_{m,{\lbh}}= 0\right) }
\sum_{\mu =1}^m
\etauh
{\mbf}_{I_\mu }^{(m)}({\br %
}_{I_1},{\br}_{I_2},{\br}_{I_3},\cdots ,{\br}_{I_m})
\nonumber \\
&=&\frac 1{N_{{\lbh}}}
\frac 1{m!}
\sum_{\left\{ I_1,I_2,\cdots ,I_m\right\} }^{\left( I_{1,{\lbh}%
},I_{2,{\lbh}},\cdots ,I_{m,{\lbh}}= 0\right) }
m
\etanh
{\mbf}_{I_1 }^{(m)}({\br %
}_{I_1},{\br}_{I_2},{\br}_{I_3},\cdots ,{\br}_{I_m})
\nonumber \\
&=&
\frac {1}{\left( m-1 \right )!}
\sum_{i_1=1}^{n}
\sum_{\left\{ I_2,I_3,\cdots ,I_m \right\} }^{\left( I_{2,{\lbh}%
},I_{3,{\lbh}},\cdots ,I_{m,{\lbh}}= 0\right) }
\etanh
{\mbf}_{i_1 }^{(m)}({\br %
}_{i_1},{\br}_{I_2},{\br}_{I_3},\cdots ,{\br}_{I_m}). \label{et03.c.6}
\end{eqnarray}

Equation (\ref{et03.c.3}) is valid for
$s^\prime = s-1 = 0, 1, 2, \cdots, m-2$ in the above case,
while Eq. (\ref{et03.c.6})
is essentially the situation with $s^\prime = s-1 =m-1$, namely
\begin{equation}
{\mbF}_{c3,{\lbh}}^{(m)}={\mbF}_{c2,s=m,{\lbh}}^{(m)}.
   \label{et03.c.7}
\end{equation}
For fixed ${\br}_{i_1}$, only $s^\prime = 0, 1, 2, \cdots, m-1$
of the remaining $m-1$ ions in the $R_{\lbh}$ part can appear
in the $T_{\lbh}=0$ slab and the remaining ion(s) in
the $R_{\lbh}^\prime$ part, then:

\begin{eqnarray}
&&
\frac 1{ \left(m-1\right) !}
\sum_{\left\{ I_2,I_3,\cdots ,I_m \right\} }^{\left( I_{2,{\lbh}%
},I_{3,{\lbh}},\cdots ,I_{m,{\lbh}} \geq 0\right) }
{\mbf}_{i_1 }^{(m)}({\br %
}_{i_1},{\br}_{I_2},{\br}_{I_3},\cdots ,{\br}_{I_m})
\nonumber \\ \label{et03.c.8}
&=&
\sum_{s^\prime=0}^{m-1}
\frac 1{ (s-1)!\left(m-s\right) !}
\sum_{\left\{ I_2,I_3,\cdots ,I_s \right\} }^{\left( I_{2,{\lbh}%
},I_{3,{\lbh}},\cdots ,I_{s,{\lbh}}= 0\right) }
\sum_{\left\{
I_{s+1},I_{s+2},\cdots ,I_m\right\} }^{\left( I_{s+1,{\lbh}},
I_{s+2,{\lbh}%
},\cdots ,I_{m,{\lbh}}>0\right) }
{\mbf}_{i_1 }^{(m)}({\br %
}_{i_1},{\br}_{I_2},{\br}_{I_3},\cdots ,{\br}_{I_m}) .    \nonumber \\
&&
\label{et03.c.8}
\end{eqnarray}
Furthermore
\begin{eqnarray}
{\mbF}_{c2+c3,{\lbh}}^{(m)}&=&
{\mbF}_{c3,{\lbh}}^{(m)}+
\sum_{s^\prime=0}^{m-2}
{\mbF}_{c2,s,{\lbh}}^{(m)}
=
\sum_{s^\prime=0}^{m-1}
{\mbF}_{c2,s,{\lbh}}^{(m)}   \nonumber \\
&=&
\sum_{s^\prime=0}^{m-1}
\sum_{i_1=1}^{n}
\frac  {\etanh}
{ (s-1)!\left(m-s\right) !}
\sum_{\left\{ I_2,I_3,\cdots ,I_s \right\} }^{\left( I_{2,{\lbh}%
},I_{3,{\lbh}},\cdots ,I_{s,{\lbh}}= 0\right) }
\sum_{\left\{
I_{s+1},I_{s+2},\cdots ,I_m\right\} }^
{\left( I_{s+1,{\lbh}},I_{s+2,{\lbh}%
},\cdots ,I_{m,{\lbh}}>0\right) }
{\mbf}_{i_1 }^{(m)}({\br %
}_{i_1},{\br}_{I_2},{\br}_{I_3},\cdots ,{\br}_{I_m})
\nonumber \\ \label{et03.c.8}
&=&
\sum_{i_1=1}^{n}
\frac  {\etanh}
{ \left(m-1\right) !}
\sum_{\left\{ I_2,I_3,\cdots ,I_m \right\} }^{\left( I_{2,{\lbh}%
},I_{3,{\lbh}},\cdots ,I_{m,{\lbh}} \geq 0\right) }
{\mbf}_{i_1 }^{(m)}({\br %
}_{i_1},{\br}_{I_2},{\br}_{I_3},\cdots ,{\br}_{I_m}) . \label{et03.c.9}
\end{eqnarray}

Similarly, Eq. (\ref{et03.b.5}) is valid for $t = 1, 2, \cdots, m-1$
in the first case, while Eq. (\ref{et03.c.9})
is essentially the situation with $t=0$, namely
\begin{equation}
{\mbF}_{c2+c3,{\lbh}}^{(m)}={\mbF}_{c1,t=0,{\lbh}}^{(m)}.
   \label{et03.d.6}
\end{equation}
For fixed ${\br}_{i_1}$, only $t = 0, 1, 2, \cdots, m-1$ of the
remaining $m-1$ ions can appear in the $L_{\lbh}$ part with
the remaining ion(s) in the $R_{\lbh}$ part at the same time, then
\begin{eqnarray}
&&
\frac 1{ \left(m-1\right) !}
\sum_{\left\{ I_2,I_3,\cdots ,I_{m} \right\} }
{\mbf}_{i_1 }^{(m)}({\br %
}_{i_1},{\br}_{I_2},{\br}_{I_3},\cdots ,{\br}_{I_m}) \nonumber \\
&=&
\sum_{t=0}^{m-1}
\frac  1{t!\left(m-t-1\right) !}
\sum_{\left\{ I_2,I_3,\cdots ,I_{t+1}\right\} }^{\left( I_{2,{\lbh}%
},I_{3,{\lbh}},\cdots ,I_{t+1,{\lbh}}< 0\right) }
\sum_{\left\{
I_{t+2},I_{t+3},\cdots ,I_m \right\} }^{\left( I_{t+2,{\lbh}},
I_{t+3,{\lbh}%
},\cdots ,I_{m,{\lbh}} \geq 0\right) }
{\mbf}_{i_1 }^{(m)}({\br %
}_{i_1},{\br}_{I_2},{\br}_{I_3},\cdots ,{\br}_{I_m})
 .    \nonumber \\
&&\label{et03.d.7}
\end{eqnarray}

As a result,
\begin{eqnarray}
{\mbF}_{c1+c2+c3,{\lbh}}^{(m)}&=&
{\mbF}_{c2+c3,{\lbh}}^{(m)}+
\sum_{t=1}^{m-1}
{\mbF}_{c1,t,{\lbh}}^{(m)}
=
\sum_{t=0}^{m-1}
{\mbF}_{c1,t,{\lbh}}^{(m)}
 \nonumber \\
&=&
\sum_{t=0}^{m-1}
\sum_{i_1=1}^n
\frac  \etanh
{ t!\left(m-t-1\right) !}
\sum_{\left\{ I_2,I_3,\cdots ,I_{t+1}\right\} }^{\left( I_{2,{\lbh}%
},I_{3,{\lbh}},\cdots ,I_{t+1,{\lbh}}< 0\right) }
\sum_{\left\{
I_{t+2},I_{t+3},\cdots ,I_m \right\} }^
{\left( I_{t+2,{\lbh}},I_{t+3,{\lbh}%
},\cdots ,I_{m,{\lbh}} \geq 0\right) }
{\mbf}_{i_1 }^{(m)}({\br %
}_{i_1},{\br}_{I_2},{\br}_{I_3},\cdots ,{\br}_{I_m})  \nonumber \\
&=&
\sum_{i_1=1}^n
\frac  \etanh
{ \left(m-1\right) !}
\sum_{\left\{ I_2,I_3,\cdots ,I_{m} \right\} }
{\mbf}_{i_1 }^{(m)}({\br %
}_{i_1},{\br}_{I_2},{\br}_{I_3},\cdots ,{\br}_{I_m})   \nonumber \\
&=&
\sum_{i_1=1}^n
\frac  {\left(\bhz-\br_{i_1} \right) \cdot \sh}{ {\Omega}}
{\mbF}_{i_1 }^{(m)}
 , \label{et03.d.9}
\end{eqnarray}
where  as defined in Eq. (\ref{et02.4.2}),
${\mbF}_{i_1 }^{(m)}$ is the net $m$-body force acting on
MD ion $i_1$ by all other $m-1$ ions in all possible configurations.

By using Eq. (\ref{et02.4}),  Eq. (\ref{et03.d.9}) can be reduced as
\begin{equation}
{\mbF}_{c1+c2+c3,{\lbh}}^{(m)}=
\sum_{i_1=1}^n
\frac  {\left(-\br_{i_1} \right) \cdot \sh}{ {\Omega}}
{\mbF}_{i_1 }^{(m)}
=
-\frac 1{ {\Omega}}
\sum_{i_1=1}^n
\left( {\mbF}_{i_1 }^{(m)}  \otimes \br_{i_1} \right)  \cdot \sh
 . \label{et03.d.94}
\end{equation}
Then the averaged net force acting on the half-line-cell
bar $B_{\lbh}$ by the $L_{\lbh}$ part in Fig. 1 is
\begin{equation}
\overline\mbF_{\lbh}^\prime
= \mbF_{\lbh} - \sum_{m=2}^M
{\mbF}_{c1+c2+c3,{\lbh}}^{(m)}
= \mbF_{\lbh} + \frac 1\Omega \sum_{i_1=1}^n
\left( \mbF_{i_1} \otimes \br_{i_1} \right) \cdot \sh,
\end{equation}
where Eq. (\ref{et01.000}) is used.
Now, let us introduce another tensor
\begin{equation}
\Tep_p=\frac 1\Omega \sum_{i_1=1}^n\mbF_{i_1}
\otimes \br_{i_1}, \label{i04.2}
\end{equation}
which is zero when an equilibrium state is reached. This provides
\begin{equation}
\overline\mbF_{\lbh}^\prime = \mbF_{\lbh} +\Tep_p\cdot\sh=
\left(\Tepmain + \Tep_p\right)\cdot\sh=\Tep\cdot\sh,\label{newi04}
\end{equation}
where the full interaction term of the internal stress is
\begin{equation}
\Tep=\Tepmain+\Tep_p.\label{i04}
\end{equation}
It can also be written as
\begin{equation}
\Tep=-\frac 1\Omega \sum_{\bz\in\text{DOF}}
\left( \frac{\partial E_{p,cell}}{\partial\bz}\right) \otimes \bz,
\label{verynew}
\end{equation}
where DOF refers to all degrees of freedom of the system
including the three period vectors $\ba$, $\bb$, $\bc$, and
all MD ion position vectors $\br_{1}$, $\br_{2}$, $\cdots$,
and $\br_{n}$. Then the period dynamics Eq. (\ref{b21.9})
can be updated into
\begin{equation}
\alpha _{\lbh,\lbh} \ddot\bh
= \left( \Tep +\Tup\right ) \cdot \sh\ \ (\bh=\ba, \bb, \bc).
\label{nb222pp}
\end{equation}

\section{Momentum Transportation}  \label{sec:transportmomentum}

Consider an ideal gas in a fixed and closed container in an
equilibrium state from a macroscopic point of view and
imagine to cut it into a left half and a right half. Gas particles
carrying their momentum can freely run between the two
halves. Then we have two choices to study it.

One choice is to employ a material-based system. In such a
system definition, the gas particles always belong to the same
half system, which they belong to at the very beginning. Then at
the very beginning, we have very clear half systems of gas particles.
However very soon, some gas particles in one half may move into
the other half, but still belong to the original half system. Then the
half systems would no longer have a clear boundary. Definitely,
Newton's second law still applies to the half systems, 
but not easy to use.

The other choice is to employ a space-based system
definition\cite{malvern}, in which at any time, a particle
belongs to the system if it is inside the 
corresponding space with  
a fixed and close geometric boundary, 
otherwise it is not. Then each half system is actually defined 
by the corresponding fixed half space inside the container, 
then always has a clear boundary. When a gas
particle moves from one half into the other half, 
it leaves from the former system and joines the
later system. The later system gets its momentum and the
former system gets its momentum as well but in the opposite
direction. 
For the dynamical process of each space-based half system, 
the total 
regular force we see is the net external force acting on the 
gas particles by the container during collisions between 
them, which is not zero at non-zero absolute temperature.  
However, the total momentum of each 
half system is always zero. Then 
the net momentum transported into and out of the half
system per unit time, due to gas particles' crossing the
boundary between the two halves, should also be considered
as an external force acting on the half system in order to
satisfy Newton's second law. 
As a matter of fact, these two
forces balance each other. Let us call the later as the force
associated with momentum transportation. Since a specific
momentum transported from space-based system $S_A$ to
its neighbour space-based system $S_B$ per unit time
should be regarded as an external force acting on
system $S_B$ by system $S_A$, its opposite direction 
momentum movement rate from  
$S_B$ to $S_A$
should be regarded as another external force acting 
on system $S_A$ by system $S_B$. They are actually
action and re-action forces satisfying Newton's third law.

Similarly, if a material-based system is used to study
each half part of the crystal above, when an ion 
``runs from one part into the other part of the crystal", 
it still
belongs to the original part, then the motion of
every individual ion must be traced all the time.
Furthermore the corresponding components of the
external stress $\Tup$, acting on the surface of one
part of the crystal, should also be identified acting
on the ions, which belong to the other part. If a
space-based system is employed, these are not
needed, but the force associated with momentum
transportation should be considered as an external
force on the system.

As in our previous work\cite{lgcjp1}, 
for the $L_{\lbh}$ and $R_{\lbh}$ parts in Fig. 1,
both as space-based systems, consider the above
statistics over the indistinguishable translated states
with the help of plane $Q_{\lbh}Q_{\lbh}^\prime$
again, but of the force associated with momentum
transportation.
If the total amount of such indistinguishable translated
states is assumed as the cell volume $\Omega$, the
amount of those where MD ion $i$ can cross
plane $Q_{\lbh}Q_{\lbh}^\prime$ during a
unit time is $\left| \dot\br_i \cdot \sh\right|$, with
momentum $m_i \dot\br_i$ being carried each.
Then the additional averaged force associated with
momentum transportation on $R_{\lbh}$ part
\begin{equation}
\mbf_{\lbh,tm}=\frac 1\Omega
\sum_{i=1}^n \left( \dot \br_i \cdot \sh \right) m_i\dot \br_i
=\frac 1\Omega \sum_{i=1}^n m_i
\left( \dot \br_i \otimes \dot \br_i \right) \cdot \sh
\end{equation}
should be added to $\overline\mbF_{\lbh}^\prime$.
As a result, Eq. (\ref{newi04}) is updated to
\begin{equation}
\overline\mbF_{\lbh}=\overline\mbF_{\lbh}^\prime +
\mbf_{\lbh,tm} = \left(\Tep + \Ttau^\prime\right)\cdot\sh,
\label{newi05}
\end{equation}
where the instantaneous kinetic-energy term of the
internal stress is
\begin{equation}
\Ttau^\prime=\frac 1\Omega
\sum_{i=1}^n m_i\dot \br_i \otimes \dot \br_i .
\label{newi06}
\end{equation}
Defining the instantaneous internal stress  as
\begin{equation}
\Tpi^\prime=\Tep+\Ttau^\prime,  \label{nnnn}
\end{equation}
the period dynamics Eq. (\ref{nb222pp}) becomes
\begin{equation}
\alpha _{\lbh,\lbh} \ddot\bh=\left( \Tpi^\prime +
\Tup\right ) \cdot \sh\ \ (\bh=\ba, \bb, \bc).\label{newb22}
\end{equation}

The observable period vectors showing fixed values
under certain external conditions (e.g. constant
external pressure and temperature)
should not depend on the directions of ions' motions.
A further unweighted average of
Eq. (\ref{newb22}) was performed over all moving
directions of the MD ions.  
For this, the averaged Eq. (\ref{newi06})
becomes:
\begin{equation}
\overline{\Ttau^\prime}=\frac 1{3\Omega}\sum_{i=1}^n m_i\left|
\dot \br_i\right| ^2\TI=\frac 2{3\Omega}E_{k,MD,ion}\TI,
\end{equation}
where $E_{k,MD,ion}$ is the total kinetic-energy of
the MD ions. Also considering the motion of 
the valence electrons the same way, 
the averaged kinetic-energy
term of internal stress should be:
\begin{equation}
\Ttau=\overline{\Ttau^\prime} + \frac 2{3\Omega}E_{k,MD,ve}\TI = 
\frac 2{3\Omega}\left(E_{k,MD,ion} + E_{k,MD,ve}\right)\TI,
\end{equation}
where $E_{k,MD,ve}$ is the total kinetic-energy of 
the valence electrons in the MD cell.
The forces corresponding to this part of the internal
stress should be balanced by the part of the
external forces involved in collisions between the ions
in the bulk surface and the surrounding external walls,
as in the above example of an ideal gas. Accordingly,
the averaged internal stress from Eq. (\ref{nnnn}) is
\begin{equation}
\Tpi=\Tep+\Ttau.\label{newc09.22}
\end{equation}
Then the period dynamics Eq. (\ref{newb22}) changes into
\begin{equation}
\alpha _{\lbh,\lbh} \ddot\bh = \left( \Tpi +\Tup\right )
\cdot \sh\ \ (\bh=\ba, \bb, \bc).\label{nb222}
\end{equation}

\section{Summary and Discussion}\label{sec:Summary}

Keeping Newton's second law for MD ions and applying it
to macroscopic half-systems with additional statistics
over indistinguishable translated states and forces associated
with momentum transportation applied, we arrived at the
coupled dynamical equations, Eq. (\ref{b03}) for MD ions
and Eq. (\ref{nb222}) for the period vectors, of crystals of
many-body interactions under constant external stress.
Equation (\ref{nb222}) shows that the system period vectors are
driven by the imbalance between the internal and external
stresses. 
Then when the system reaches an equilibrium state, 
the internal and external
stresses balance each other. 
The internal stress has both full kinetic-energy and full
interaction terms. As a result,  the dynamical equations
and associated formulas in this article for many-body
interactions share the same form of those in our last
work\cite{lgcjp1} for pair-potential only.

The kinetic-energy term was obtained from the
statistics of forces associated with momentum
transportation when the two halves of the system
are recognized as space-based ones.
Since the full interaction term of the internal stress
Eq. (\ref{verynew}) is valid for any-body interactions,  
it should also be valid for forces from electrons
but calculated based on quantum mechanics
involved. In such a situation, the effective interactions 
among ions through electrons are many-body ones, 
as the calculated state of electrons depends on the 
positions of all ions.

As a matter of fact, the external stress is required as
a constant only in deriving Eqs. (\ref{b02})
and (\ref{b04}) and this  requirement only means
that it is constant over the surface of the crystal
throughout this paper. Then for such external
stress but changing with real time, one can solve
the crystal with Eqs. (\ref{b03}) and (\ref{nb222})
to reach an equilibrium state iteratively for the given
external stress at a given real time, then the next
real time, ..., till end. 

In the MD world, simulations are usually classified
into various ensembles, based on applicable
combinations of fixed volume, constant external
pressure, and constant external temperature.
For ensembles of fixed volume, Eq. (\ref{nb222})
shows that an external stress balancing the
internal stress should always be supplied or
assumed. For ensembles of constant external
pressure/stress, Eq. (\ref{nb222}) can be used.
Additionally, the straightforward ion speed
rescaling method can always be a choice, for
constant external temperature simulations.

\begin{acknowledgments}
The author wishes to thank Prof. Ding-Sheng Wang,
Institute of Physics, Prof. Si-Yuan Zhang, Changchun
Institute of Applied Chemistry, Prof. S. S. Wu, Jilin
University, Prof. En-Ge Wang, Beijing University,
P.R. China, Dr. Kenneth Edgecombe,
Dr. Hartmut Schmider, Dr. Malcolm J. Stott,
Dr. Kevin Robbie, Queen's University, Canada,
Dr. Xiaohua Wu, the Royal Military College of Canada,
and Biomolecular Simulation Team, Compute Canada,
for their helpful discussions and earnest encouragement,
and also the Centre for Advanced Computing,
Queen's University, Canada, for its support.
\end{acknowledgments}


\begin{references}


\bibitem{ktl}  C. Kitte, 1986, 
{\it Introduction to Solid State Physics} 
(John Wiley \& Sons, New York).


\bibitem{ibach}  H. Ibach, ~and H. L\"{u}th, 2009, 
{\it Solid-state physics: an introduction to principles 
of materials science} 
(Springer, New York).


\bibitem{patterson}  J.D. Patterson, ~and B.C. Bailey, 2010, 
{\it Solid-state physics: introduction to the theory} 
(Springer, New York).



\bibitem{a}  H.C. Andersen, 1980, J. Chem. Phys.
\textbf {72}, 2384.
doi:10.1063/1.439486.

\bibitem{pr1}  M. Parrinello,~and A. Rahman, 1980, 
Phys. Rev. Lett. \textbf {45}, 1196.
doi:10.1103/PhysRevLett.45.1196.


\bibitem{pr2}  M. Parrinello,~and A. Rahman, 1981, 
J. Appl. Phys. \textbf {52}, 7182. 
doi:10.1063/1.328693.




\bibitem{al}  M.P. Allen,~and D.J. Tildesley, 1987, 
{\it Computer Simulation of
Liquids} (Oxford University, Oxford).

\bibitem{fre}  D. Frenkel,~and B. Smit, 1996, 
{\it Understanding Molecular
Simulation} (Academic, New York).

\bibitem{haile}  J. M. Haile, 1992, 
{\it Molecular Dynamics Simulation Elementary
Methods} (John Wiley \& Sons, New York).



\bibitem{ray1}  J.R. Ray, 1983, J. Chem. Phys. \textbf {79}, 5128.
doi:10.1063/1.445636.

\bibitem{nose}  S. Nos\'{e},~and M.L. Klein, 
1983, Mol. Phys.
\textbf {50}, 1055.
doi:10.1080/00268978300102851.


\bibitem{ryckaert}
J.P. Ryckaert,~and G. Ciccotti, 
1983, J. Chem. Phys.
\textbf {78}, 7368.
doi: 10.1063/1.444728.



\bibitem{evans}
D.J. Evans,~and G.P. Morriss
1984, Computer Physics Reports
\textbf {1}, 297.
doi:10.1016/0167-7977(84)90001-7.

\bibitem{ray2}  J.R. Ray, ~and A. Rahman, 1984, 
J. Chem. Phys. \textbf {80}, 4423.
doi:10.1063/1.447221. 

\bibitem{ray3}  J.R. Ray,~and A. Rahman, 1985, J. Chem. Phys. 
\textbf {82}, 4243.
doi:10.1063/1.448813. 


\bibitem{hoover1985}
W.G. Hoover,
1985, Phys. Rev. A 
\textbf {31}, 1695.
doi:10.1103/PhysRevA.31.1695.



\bibitem{hoover1986}
W.G. Hoover,
1986, Phys. Rev. A 
\textbf {34}, 2499.
doi:10.1103/PhysRevA.34.2499.


\bibitem{cll}  C.L. Cleveland, 1988,  J. Chem. Phys. \textbf {89}, 4987.
doi:10.1063/1.455642.

\bibitem{wen}  R.M. Wentzcovitch, 1991, Phys. Rev. B \textbf {44}, 2358.
doi:10.1103/PhysRevB.44.2358.


\bibitem{melchionna}  
S. Melchionna,~G. Ciccotti,~and B.L. Holian
1993, Mol. Phys.
\textbf {78}, 533.
doi:10.1080/00268979300100371.

\bibitem{jv}  J.V. Lill,~and J.Q. Broughton, 1994, Phys. Rev. B
\textbf {49}, 11 619.
doi:10.1103/PhysRevB.49.11619.

\bibitem{fo}  P. Focher, G.L. Chiarotti, M. Bernasconi, E. Tosatti,~ and M.
Parrinello, 1994, Europhys. Lett. \textbf {26}, 345.
doi:10.1209/0295-5075/26/5/005.


\bibitem{martyna}
G.J. Martyna,~D.J. Tobias,~and M.L. Klein,
1994, J. Chem. Phys.
\textbf {101}, 4177.
doi:10.1063/1.467468.


\bibitem{ber}  M.Bernasconi,~G.L. Chiarotti, P. Focher, S. Scandolo,~E.
Tosatti,~ and M. Parrinello, 1995, J. Phys. Chem. Solids \textbf {56}, 501.
doi:10.1016/0022-3697(94)00228-2.

\bibitem{szm}  I. Souza,~and J.L. Martins, 1997, Phys. Rev. B
\textbf {55}, 8733.
doi:10.1103/PhysRevB.55.8733.


\bibitem{lww}  G. Liu, E.G. Wang,~and D.S. Wang, 1997, Chin.
Phys. Lett. \textbf {14}, 764.
doi:10.1088/0256-307X/14/10/012.

\bibitem{lwssc}  G. Liu,~and E.G. Wang, 1998, Solid State
Commun. \textbf {105}, 671.
doi:10.1016/S0038-1098(98)00001-5.


\bibitem{tuckerman}
M.E. Tuckerman,~J. Alejandre,~R. L\'{o}pez-Rend\'{o}n,~A.L. Jochim,~and G.J. Martyna
2006, J. Phys. A: Math. Gen. 
\textbf {39}, 5629.
doi:10.1088/0305-4470/39/19/S18.



\bibitem{crystalpredictions}
C.C. Pantelides,~C.S. Adjiman,~and A.V. Kazantsev, 2014, 
Top. Curr. Chem. \textbf{345}, 25.
doi:10.1007/128\_2013\_497.



\bibitem{lgcjp1}  G. Liu, 2015, Can. J. Phys. \textbf {93}, 974, 
doi:/10.1139/cjp-2014-0518.


\bibitem{aidan}
A.P. Thompson,~S.J. Plimpton,~and W. Mattson, 
2009, J. Chem. Phys.
\textbf {131}, 154107.
doi:10.1063/1.3245303.

\bibitem{malvern}
L. E. Malvern, 1969,  
{\it Introduction To The Mechanics of A
Continuous Medium} (Prentice-Hall, Englewood Cliffs,
New Jersey).


\end{references}
\end{document}